\documentclass[12pt]{iopart}

\usepackage{setstack}
\usepackage{bm}
\usepackage{graphicx}
\usepackage{amsbsy}
\usepackage{amstext}
\usepackage{amssymb}
\usepackage{units}
\usepackage{hyperref}

\begin{document}

\title{Temperature effect on the magnetic oscillations in 2D materials}	


\author{F Escudero, J S Ardenghi,
	P Jasen}

\address{IFISUR, Departamento de F\'isica (UNS-CONICET), Av. Alem 1253, B8000CPB Bah\'ia Blanca, Argentina\\
	Instituto de F\'isica del Sur (IFISUR, UNS-CONICET)\\
	Av. Alem 1253, B8000CPB Bah\'ia Blanca, Argentina}
\ead{federico.escudero@uns.edu.ar}

\begin{abstract}
	We study the magnetic oscillations (MO) in 2D materials with a buckled honeycomb lattice, considering a perpendicular electric and magnetic field. At zero temperature the MO consist of the sum of four sawtooth oscillations, with two unique frequencies and phases. The values of these frequencies depend on the Fermi energy and electric field, which in turn determine the condition for a beating phenomenon in the MO. We analyse the temperature effect in the MO by considering its local corrections over each magnetization peak, given by Fermi-Dirac like functions. We show that the width of these functions is related to the minimum temperature necessary to observe the spin and valley properties in the MO. In particular, we find that in order to observe the spin splitting, the width must be lower than the MO phase difference. Likewise, in order to observe valley mixing effects, the width must be lower than the MO period. We also show that at high temperatures, all the maxima and minima in the MO are shift to a constant value, in which case we obtain a simple expression for the MO and its envelope. The results obtained show unique features in the MO in 2D materials, given by the interplay between the valley and spin. 
\end{abstract}


\section{Introduction}

Since the experimental realization of graphene in 2004 \cite{Novoselov2005,Geim2007,Zhang2005}, many similar
planar systems have been studied \cite{Mas-Balleste2011,Lin2016,Gupta2015}. Among them are
silicene \cite{Zhao2016,Lay2015,Zhuang2015,Houssa2015}, germanene \cite{Balendhran2014,Davila2016}, stanene \cite{Saxena2016,Zhu2015} and phosphorene \cite{Carvalho2016,Cho2017}. These materials have
a 2D hexagonal lattice, made of two buckled subtallices
A and B. They are best described with a tight binding (TB) model,
which leads to an effective Dirac-like Hamiltonian in the low energy
approximation \cite{Liu2011,Spencer2016}. Thus these materials are also referred as Dirac crystals.
Despite their similarities, there are important features that distinguish
one material from another. One is the spin-orbit interaction (SOI),
which is very small in graphene (about $10^{-3}$ meV \cite{Neto2009}), but relatively large
in other materials (for instance, it is 0.1 eV in stanene \cite{Liu2011}),
which makes them a topological insulator \cite{Kou2017,Tahir2013,Ezawa2015,Huang2016}. Moreover, a strong SOI would
make possible the observation of the quantum spin Hall effect \cite{Hsu2017,Wang2016,Zhang2016,Ghazaryan2015,Ezawa2012,Liu2017}. Another
characteristic is the buckle height, which defines the layer separation
between the two sublattices. In graphene this buckle height is zero \cite{Neto2009},
but it is not zero in the other Dirac crystals \cite{Gupta2015}. Hence, applying
a perpendicular electric field between the two sublattices causes a
potential difference, which splits the energy bands and can be used
to tune the bandgap \cite{Drummond2012,Du2014,Aghaei2015,Yan2015,Ni2011,Abbasi2018,Wang2017}.

The magnetic properties of the Dirac crystals have been investigated
in recent works \cite{Sharapov2004,Tabert2015,Hese2014,Fu2011,Lukyanchuk2011}. Unlike in conventional materials, the magnetization
in these systems has unique features \cite{Uchoa2008,Ardenghi2015,Ardenghi2014,Escudero2017a,Escudero2018b,Escudero2018}. A particular interesting phenomenon are the magnetic oscillations (MO), the so called de Haas van Alphen effect \cite{Shoenberg1984}, produced by the discrete energy levels that appear when a magnetic field is applied. At zero temperature the MO are sawtooth \cite{Fu2011}, with the peaks been caused by the change in the last occupied energy level \cite{Escudero2017}. Therefore the MO depend strongly on the system energy levels. In the Dirac crystals, at low energies the
dispersion relation is relativistic \cite{Neto2009}, which causes the Landau levels (LL) to be not
equidistant \cite{Goerbig2011,Ardenghi2013}. These anomalous LL can be modify by external parallel and perpendicular electric fields \cite{Lukose2007,Goerbig2011}. For instance, the MO in graphene can be modulated by an in-plane electric field \cite{Zhang2010}, which leads to unique features not seen in the conventional 2D electron gas.
In the Dirac crystals, a perpendicular electric field alters the LL due to the buckled height and strong SOI \cite{Escudero2018}, which can have an appreciable effect in the MO.

At non zero temperature the MO are broadened as a result of the Fermi-Dirac distribution. In classical metals this is described
by the Lifshitz-Kosevich (LK) formula \cite{Shoenberg1984}, which incorporates the temperature effect as a reduction factor. This formula has been extended to the case of
Dirac crystals \cite{Sharapov2004}, where the difference only lies in the form of the reduction
factor. Another approach, recently developed in graphene \cite{Escudero2018a}, considers the temperature effect by local corrections over each MO peak. This is particular useful at very low temperatures, where the MO are modified only around the peaks location at $T=0$. Nevertheless, there is no detailed analysis about how the \textit{fine structure} of the MO in 2D buckled materials is progressively lost as the temperature increases. This is particular relevant from an experimental point of view, since there is always a limit to how low the temperature can be.

Motivated by this we analysed the MO in a general pristine Dirac crystal, in the presence of a perpendicular
electric and magnetic field, taking into account the Zeeman effect.
We have organized this work as follow: in section 2 we describe the MO at zero temperature, showing that it consist of two unique frequencies and phases. Then we study the dependence of these frequencies with the electric field, and the condition for a beating phenomenon. In section 3 we study how the temperature broadens the MO and affects its observation. We estimate the minimum temperature required to observe the valley and spin properties. At high temperature we also obtain a simple approximation for the MO and its envelope. Finally, our conclusions follow in section 4.  

\section{MO at zero temperature}

\begin{figure}[b]
	\includegraphics[scale=0.35]{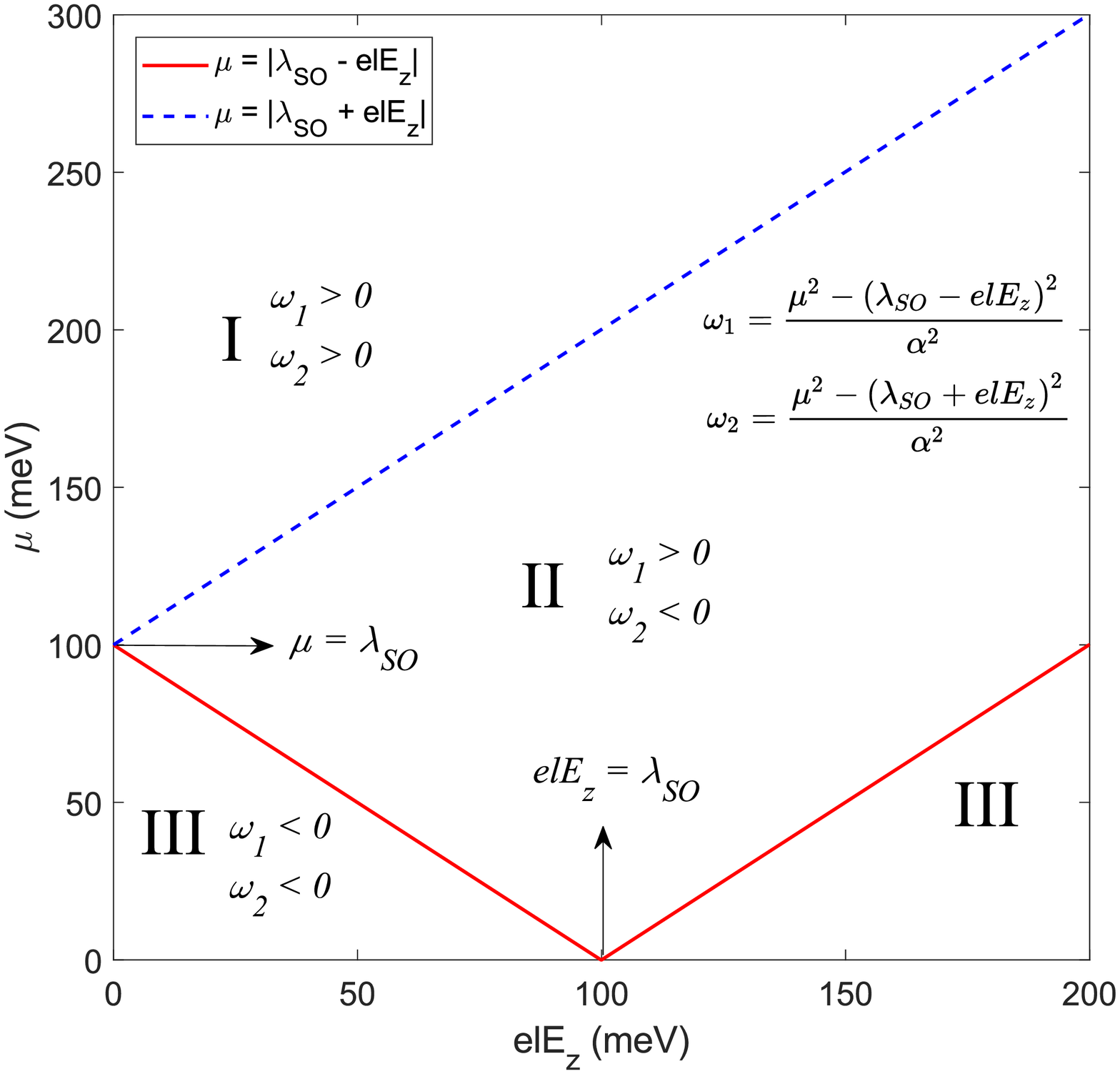}	
	\caption{MO frequency spectrum for stanene, as a function of the perpendicular
		electric field $E_{z}$ and the Fermi energy $\mu$. The presence
		of $E_{z}$ produces a broken valley degeneracy, which results in
		two frequencies $\omega_{1}$ and $\omega_{2}$ for the MO, with $\omega_{1}>\omega_{2}$
		always. The oscillations occur only if $\omega>0$, which defines
		the three regions shown: (I) $\omega_{1}>0$ and $\omega_{2}>0$,
		so both frequencies are present in the MO, (II) $\omega_{1}>0$ and
		$\omega_{2}<0$, in which case the magnetization oscillates with only one frequency, as in graphene, and (III) $\omega_{1}<0$ and $\omega_{2}<0$,
		so there is no MO.\label{fig1}}	
\end{figure}

We shall study the MO in a general 2D system with a buckled honeycomb
structure. Examples of these materials are silicene, germanene, stanene
and phosphorene. We will consider energies close to the Fermi
level, in which case one can apply the long wavelength approximation \cite{Neto2009,Liu2011}.
Then, in the presence of a perpendicular electric field $E_{z},$
these materials are described by a Dirac Hamiltonian of the form \cite{Spencer2016}

\begin{equation}
H=\upsilon_{\rm F}\left(\eta p_{x}\sigma_{x}+p_{y}\sigma_{y}\right)+\kappa_s^{\eta}\sigma_{z},\label{H1}
\end{equation}
where $\upsilon_{\rm F}$ is the Fermi velocity, $\boldsymbol{\sigma}$ are the Pauli
matrices and $\kappa_s^{\eta}=\eta s\lambda_{\rm SO}-elE_{z}$, with
$\lambda_{\rm SO}$ the spin-orbit coupling interaction (SOI) and $l$ the
buckle height. The indices $\eta$ and $s$ are the valley and spin indices, with values $1\left(-1\right)$ for the $K\left(K'\right)$ valley and spin up (down). The particular
values of $\upsilon_{\rm F}$, $\lambda_{\rm SO}$ and $l$ depend on the 2D
material. It is worth noting that graphene can be considered a special
case, with $l=0$ and $\lambda_{\rm SO}\simeq0$. In the presence of a
perpendicular magnetic field $B$, the Hamiltonian given by equation (\ref{H1}) gives
the energy levels \cite{Escudero2018} $\varepsilon_{\zeta,n,\eta,s}=\zeta\left[\left(s\lambda_{\rm SO}-\eta elE_{z}\right)^{2}+\alpha^{2}nB\right]^{1/2}-s\mu_{\rm B}B$,
where $\zeta=\pm1$ for the conduction and valence bands, $\alpha=\upsilon_{\rm F}\sqrt{2\hbar e}$, $n=0,\:1,\,2,\ldots$
for the Landau level (LL) and we took into account the Zeeman term $\mu_{\rm B}B$.
Each energy level has a degeneracy given by $D=\mathcal{A}B/\phi$,
where $\mathcal{A}$ is the sheet area and $\phi=h/e$ is the magnetic
unit flux \cite{Goerbig2011}. We will take a constant Fermi energy $\mu>0$, so that
only the conduction band contributes to the MO. Then the problem becomes
analogue to the one already studied in graphene \cite{Escudero2018a}, with the inclusion of the term $\kappa_s^{\eta}$.
In this way, generalizing this approach we get that the MO
are given by (see the Appendix A for details)

\begin{equation}
M=\sum_{i=1,2}\frac{A_{i}}{\pi}\sum_{s=\pm1}\arctan\left\{ \cot\left[\pi\omega_{i}\left(\frac{1}{B}+s\Delta_{i}\right)\right]\right\} ,\label{MO}
\end{equation}
where

\begin{eqnarray}
A_{i} & = & -\frac{e}{2h}\frac{\alpha^{2}\omega_{i}}{\mu},\label{Amp}\\
\omega_{i} & = & \frac{\mu^{2}-\left[\lambda_{SO}+\left(-1\right)^{i}elE_{z}\right]^{2}}{\alpha^{2}},\label{Freq}\\
\Delta_{i} & = & \frac{2\mu\mu_{\rm B}}{\alpha^{2}\omega_{i}}.\label{Delta}
\end{eqnarray}
Therefore the MO at zero temperature consist of four type of peaks,
corresponding to the possible combinations of valley and spin. There
are two unique frequencies $\omega_1$ and $\omega_2$, with phases $\Delta_1$ and $\Delta_2$. This result
generalizes the graphene case, and it says that the broken valley degeneracy
in buckled 2D materials is seen in the MO as two oscillations with
different frequency \cite{Tabert2015}. The values of these frequencies and phases
depend on the properties of the Dirac crystal, such as the SOI, the
buckle height and Fermi velocity, as well as the Fermi energy and
the perpendicular electric field. Therefore, these parameters define
the conditions for which the peaks can occur, for that implies $\omega>0$.
In graphene, the buckle height is zero and the SOI negligible, so
there is only one frequency $\omega_{g}=\omega_{1}=\omega_{2}=\mu^{2}/\alpha^{2}$
and two peaks with phase difference between them $\Delta_{g}=2\mu\mu_{\rm B}\omega_{g}\alpha^{2}$;
the condition $\omega_{g}>0$ just implies $\mu>0$. For the other
crystals, the condition $\omega>0$ implies $\mu^{2}>\left(\lambda_{\rm SO}\pm elE_{z}\right)^{2}$,
so we have the regions indicated in figure \ref{fig1}, corresponding to stanene. Depending on the value
of the Fermi energy and the electric field, three possibilities
can occur: (I) $\omega_{1}>0$ and $\omega_{2}>0$, so all 4 peaks
are present; (II) $\omega_{1}>0$ and $\omega_{2}<0$, so only two
peaks with frequency $\omega_{1}$ and phase difference $\Delta_{1}$
are present; (III) $\omega_{1}<0$ and $\omega_{2}<0$, so there are no
peaks and therefore no MO (the magnetization would be given only by
the regular, non-oscillatory contribution). Notice that $\omega_{2}$
always decrease with increasing $E_{z}$, while $\omega_{1}$ increases
with $E_{z}$ for $elE_{z}<\lambda_{\rm SO}$, it takes its maximum at
$elE_{z}=\lambda_{\rm SO}$ (where $\omega_{1}=\mu^{2}/\alpha^{2}$ as
in graphene), and then decrease with increasing $E_{z}$ for $elE_{z}>\lambda_{\rm SO}$. 
It should be noted that the MO given by equation (\ref{MO}) equals the
total magnetization only when $\mu>\left|\lambda_{\rm SO}\pm elE_{z}\right|$
\cite{Sharapov2004}, which implies $\omega_{i}>0$.
Thus only when both $\omega_{1}$ and $\omega_{2}$ are present (region
I in figure \ref{fig1}), the total magnetization is given by equation (\ref{MO}).
In the other regions, one has also to consider the regular and vacuum contributions to the total magnetization. 

The relationship between the MO and $\omega>0$ can be better understand
by analysing the energy level change that produce the oscillation.
First of all, for a constant $\mu$ we have the energy levels $\varepsilon_{i}=\left[\mu^{2}+\alpha^{2}\left(nB-\omega_{i}\right)\right]^{1/2}-s\mu_{\rm B}B$
associated with $\omega_{i}$, given by equation (\ref{Freq}). Then
$\omega_{i}<0$ implies $\left(\varepsilon_{i}+s\mu_{\rm B}B\right)^{2}>\mu^{2}+n\alpha^{2}B$,
but the occupied energy levels satisfy $\varepsilon_{i}<\mu$. Hence,
given that in general $\mu_{\rm B}B/\mu\ll1$, for $B>0$ (maintaining
the magnetic field direction), we have that $\varepsilon_{i}$ is
never occupied if $\omega_{i}<0$, so there is no oscillation associated
with a change of $\varepsilon_{i}$. In the particular case (II) in
figure \ref{fig1}, we have $\left(\lambda_{SO}-elE_{z}\right)^{2}<\mu^{2}<\left(\lambda_{SO}+elE_{z}\right)^{2}$,
and the last LL $n$ occupied in $\varepsilon_{1}$ satisfy $\left(\lambda_{SO}-elE_{z}\right)^{2}+n\alpha^{2}B<\left(\lambda_{SO}+elE_{z}\right)^{2}<\left(\lambda_{SO}-elE_{z}\right)^{2}+\left(n+1\right)\alpha^{2}B$
. Thus, when $\omega_{1}>0$ and $\omega_{2}<0$, $\varepsilon_{2}$
is not occupied and there are $n=\mathrm{Floor}\left[\left(\omega_{1}-\omega_{2}\right)/B\right]$
LL occupied in $\varepsilon_{1}.$ We see that $n$ depends on $B,$
and its value is given by the ratio between the frequency difference
and the magnetic field. This is expected considering that the magnetization
oscillates as a function of $1/B$.

\begin{figure}[b]
	\includegraphics[scale=0.35]{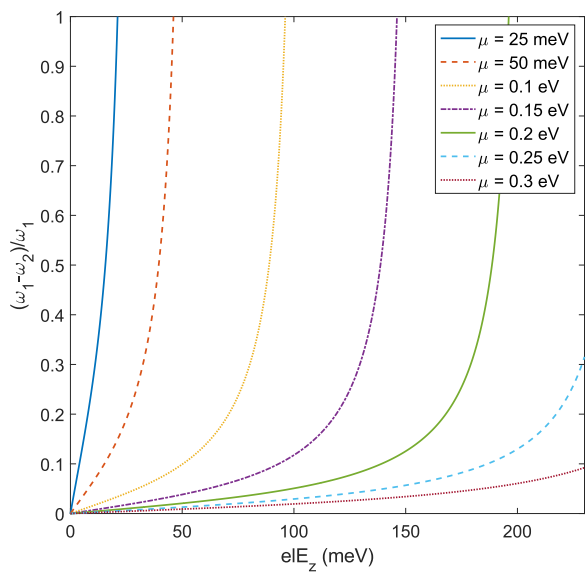}
	\caption{For silicene, plot of $\left(\omega_{1}-\omega_{2}\right)/\omega_{1}$
		as a function of the perpendicular electric field $E_{z}$, for different
		Fermi energies $\mu$. The MO frequencies $\omega_1$ and $\omega_2$ are given by equation \ref{Freq}. A beating phenomenon is observed only
		when $\left(\omega_{1}-\omega_{2}\right)/\omega_{1}\ll1$.\label{fig2}}	
\end{figure}

In the general case, when both frequencies are present, the MO will
show an interference pattern, produced by the superposition of $M_{1}$
and $M_{2}$, each one being a sawtooth oscillation. The specific
pattern in the MO will, in general, depend on the values of $\omega_{1}$
and $\omega_{2}$. The most interesting situation occurs when $\omega_{1}$
and $\omega_{2}$ are close, in which case the MO show a beating
phenomenon. Given that always $\omega_{1}>\omega_{2}$, the beating condition
is $\left(\omega_{1}-\omega_{2}\right)/\omega_{1}\ll1$. In figure \ref{fig2} it is plotted $\left(\omega_{1}-\omega_{2}\right)/\omega_{1}$ for silicene,
at different values of $\mu$, as a function of the perpendicular
electric field.
\begin{figure}[b]
	\includegraphics[scale=0.34]{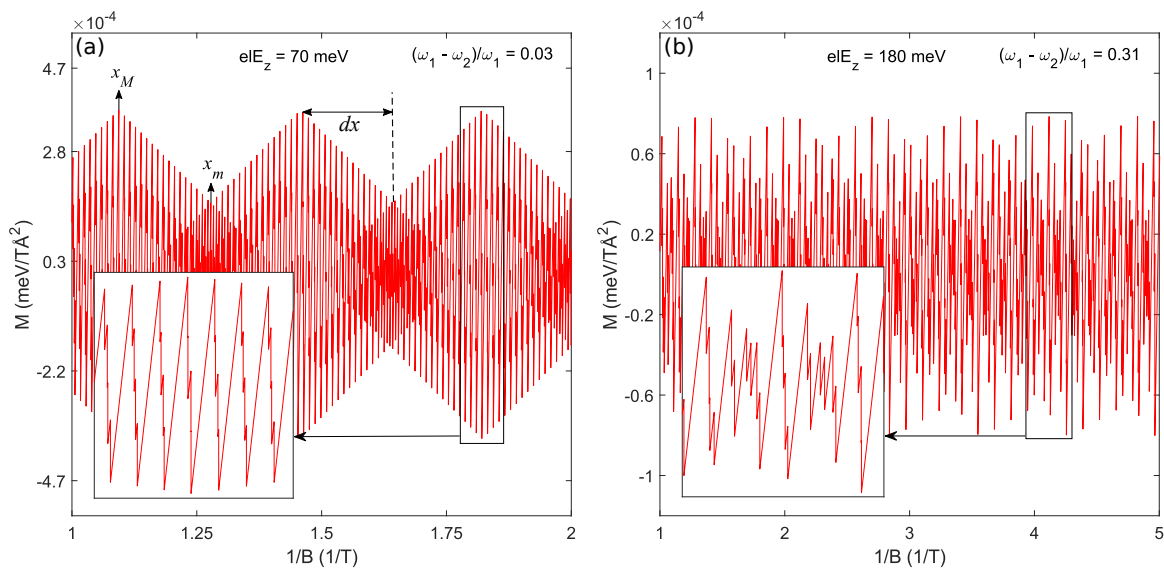}
	\caption{MO in silicene at zero temperature, with a Fermi energy $\mu=0.2$ eV, for (a)
		$\left(\omega_{1}-\omega_{2}\right)/\omega_{1}=0.03$ and (b) $\left(\omega_{1}-\omega_{2}\right)/\omega_{1}=0.31$.
		In (a) we see a clear beating phenomenon, consistent with the condition
		$\left(\omega_{1}-\omega_{2}\right)/\omega_{1}\ll1$. The beating
		is seen as a rombo-like pattern, produced by the superposition of
		the sawtooth oscillations with frequencies $\omega_{1}$ and $\omega_{2}$.
		The beating maxima and minima are $x_{M}=r/\left(\omega_{1}-\omega_{2}\right)$
		and $x_{m}=\left(r+1/2\right)/\left(\omega_{1}-\omega_{2}\right)$,
		with $r$ an integer, and their distance is $dx=1/2\left(\omega_{1}-\omega_{2}\right)=\hbar\upsilon_{\rm F}^{2}/4\lambda_{\rm SO}lE_{z}$.
		In (b), the beating condition is not satisfied, resulting in a more
		random-like pattern in the MO. In this situation, the order of the
		peaks depends strongly not only on the value of $E_{z}$, but also
		on the magnetic field $B$.\label{fig3}}
\end{figure}
We observe that the lower the Fermi energy, the lower the range of
electric field for which there is beating. In the practice,
a clear beating phenomenon is observed as long as $\left(\omega_{1}-\omega_{2}\right)/\omega_{1}\lesssim0.1$.
When $\left(\omega_{1}-\omega_{2}\right)/\omega_{1}\ll1$ is not satisfied, there is still 
an interference in the MO, only that it would not be seen as a beating
phenomenon. Instead the MO show a more random-like pattern, where
the behaviour depends specifically on the particular values of $\omega_{1}$
and $\omega_{2}$. This can be seen in figure \ref{fig3}, where we plotted the
MO in silicene for $\mu=0.2$ eV, at $elE_{z}=70$ meV and $elE_{z}=180$ meV.
This drastic change of behaviour in the MO can be explained by analysing
how the energies levels are sorted in each case. When there is beating,
$\omega_{1}$ and $\omega_{2}$ are close and so are the respective
energy levels $\varepsilon_{i,n}$ (with the same LL $n$) that give rise to these frequencies,
as discussed above. Thus the change of the last energy level (which
produces the MO) follows an ordered pattern that interchanges $\varepsilon_{1,n}$
and $\varepsilon_{2,n}$ as $B$ is changed. On the other hand, when
$\omega_{1}$ and $\omega_{2}$ are far apart, such that $\left(\omega_{1}-\omega_{2}\right)/\omega_{1}\ll1$
is not satisfied, then the energy levels $\varepsilon_{1,n}$ and
$\varepsilon_{2,n}$ are not close and there is no clear pattern in
the change of the last energy level. In this region, the sort of the
energy levels depends on the value of $E_{z}$ and $B$, leading to
a seemingly random pattern in the MO. However, it should be noted
that if one analyses the specific order of the energy levels at a particular
$E_{z}$, then the MO behaviour can be explained \cite{Escudero2018}. 

We shall now analyse in more detail the beating phenomenon in the
MO at zero temperature. As we can see in figure \ref{fig3}(a), the beating has
a rombo-like pattern, caused by both $M_{1}$ and $M_{2}$ being a
sawtooth oscillation. If we restrict to few values of magnetic
field, one can appreciate the \emph{fine structure} of the MO, as
can be seen in the zoomed area. In this region one explicitly sees
the four MO peaks given by equation (\ref{MO}), with their amplitude and
phase being determined by the change of spin and valley in the last
energy level. The absolute maxima $1/B_{M}\equiv x_{M}$ in the MO
occur when there is constructive interference. From equations (\ref{MO}),
(\ref{Freq}) and (\ref{Delta}), this implies $x_{M}=m_{1}/\omega_{1}\pm\Delta_{1}=m_{2}/\omega_{2}\pm\Delta_{2},$
where $m_{1}$ and $m_{2}$ are integer such that $m_{2}=m_{1}+r$, with $r$ an integer.
Given that in general $\omega\Delta\ll1,$ we get that the maxima
occur at $x_{M}=r/\left(\omega_{1}-\omega_{2}\right)$. The absolute minima happen between two maxima, so $x_{m}=\left(r+1/2\right)/\left(\omega_{1}-\omega_{2}\right)$,
and the width between maxima and minima is $dx=1/2\left(\omega_{1}-\omega_{2}\right)=\hbar\upsilon_{\rm F}^{2}/4\lambda_{\rm SO}lE_{z}$.
Therefore, one can obtain information about the material parameters
by measuring the width of the rombo-like pattern in the MO. Notice
that $\left(\omega_{1}-\omega_{2}\right)$ does not depend on the
magnetic field or Fermi energy, but only on the electric field. This
is expected because the width depends on the frequency difference of
the peaks, which is only produced by the perpendicular electric field.

\section{MO at non zero temperature}

We shall now study the temperature influence in the MO. We will consider low $B$ and $T$, such that we can neglect the effect of lattice vibrations \cite{Khalid1988,Engelsberg1970}.
The temperature effect in the MO can be taken into account
in different ways. The most common one is by the Lifshitz--Kosevich
(LK) formula \cite{Shoenberg1984}, in which the damping effects such as the temperature
are considered by reductions factors. Thus, in the pristine case (i.e.
no impurities), the MO at $T\ne0$ are \cite{Sharapov2004}

\begin{equation}
M_{T}=\sum_{i=1,2}A_{i}\sum_{s=\pm1}\sum_{p=1}^{\infty}\frac{R_{T}}{\pi p}\sin\left[2\pi p\omega_{i}\left(\frac{1}{B}+s\Delta_{i}\right)\right],\label{MT LK}
\end{equation}
where $R_{T}=\lambda p/\sinh\left(\lambda p\right)$ with $\lambda=4\pi^{2}\mu k_{\rm B}T/\alpha^{2}B$
. Another way to express the MO at non zero temperature is by
considering the local corrections to each peak due the Fermi-Dirac
distribution, as has been done for graphene \cite{Escudero2018a}. Generalizing
this result we obtain (see the Appendix B for details)

\begin{equation}
M_{T}=\sum_{i=1,2}\frac{A_{i}}{\pi}\sum_{s=\pm1}\arctan\left\{ \cot\left[\pi\omega_{i}\left(\frac{1}{B}+s\Delta_{i}\right)+\sum_{n}\pi\mathcal{F}_{i,n}\right]\right\} ,\label{MOT}
\end{equation}
where $\mathcal{F}_{i,n}=\left\{ 1+\exp\left[\beta\mu_{\rm B}\left(B_{n}-B\right)/B_{n}\Delta_{i}\right]\right\} ^{-1}$,
with $\beta=1/k_{\rm B}T$ and $B_{n}^{-1}=n/\omega_{i}-s\Delta_{i}$ being the MO peaks location
at $T=0$. 
\begin{figure}[b]
	\includegraphics[scale=0.55]{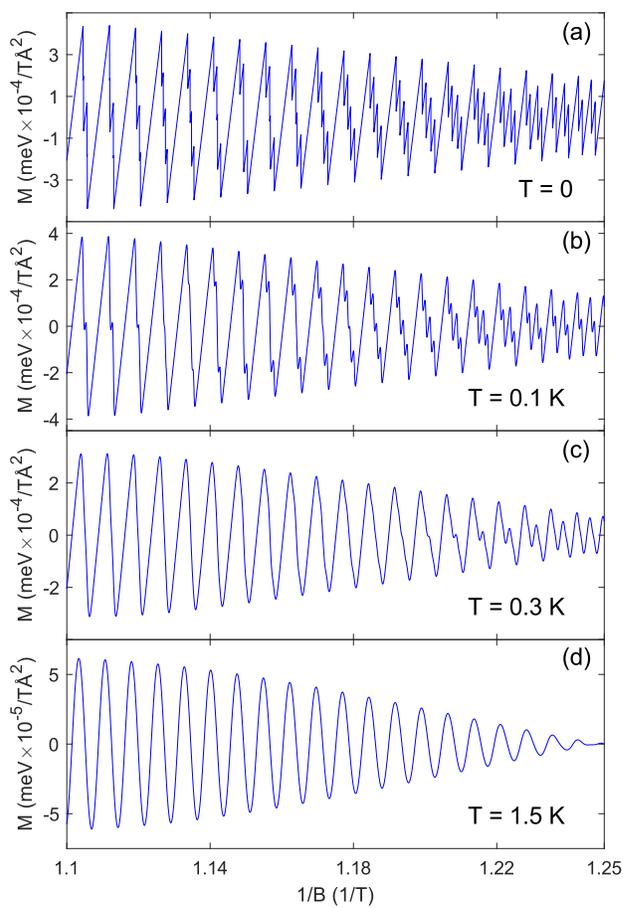}
	\caption{MO in silicene, for different temperatures, with $\mu=0.25$ eV and
		$elE_{z}=92$ meV, resulting in a beating phenomenon with $\left(\omega_{1}-\omega_{2}\right)/\omega_{1}\sim0.03$.
		All cases shown correspond to the region between the beating maximum at $x_{M}=4/\left(\omega_{1}-\omega_{2}\right)$
		and the minimum at $x_{m}=4.5/\left(\omega_{1}-\omega_{2}\right)$.
		In (a) we see the MO at $T=0$, where we can observe the \emph{fine structure}
		of the MO, due to the spin splitting (SP) and valley mixing (VM).
		For $T=0.1$ K in (b), the MO are now damped, but one can still observe
		the SP. Increasing further the temperature to $T=0.3$ K in (c), the
		small peaks due to the SP disappear, although the VM around the beating
		minimum still can be seen. If we increase further the temperature
		to $T=1.5$ K in (d), then the VM also disappears in the MO. Thus
		we say that at this stage the \emph{fine structure} of the MO is damped
		due to the temperature.\label{fig4}}
\end{figure}
It is
instructive to compare these two expressions for the MO at $T\neq0$,
because depending on the situation it may be convenient to use one or
the other formula. It is important to note that both expressions
give the same MO at $T\ne0$; they are just two different ways of
expressing the same. The series given by equation (\ref{MT LK})
express the MO as a sum of harmonics, which in some situations could
be useful, in particular when one can isolate the contribution from
each harmonic. At low temperature the summation cannot be exactly solved,
although it should be noted that few terms are needed in the infinite
sum, since the factors decay rapidly with $p$. At \emph{high }temperatures,
such that $\sinh\left(\lambda p\right)\sim\exp\left(-\lambda p\right)$,
the summation can be solved, leading to a simple expression
for the MO. On the other hand, the expression given by equation (\ref{MOT})
considers the temperature influence by local corrections around each
MO peak at $T=0$. Indeed, each term inside the summation over $n$
is a Fermi-Dirac like function, which at low temperature is appreciable
only around $B_{n}$. This is particular useful to analyse in detail
how the increase in the temperature affects the observation of the MO \emph{fine
	structure}, such as the spin splitting and valley mixing.

In figure \ref{fig4} we show the MO for different temperatures, in the case of
silicene with $\mu=0.25$ eV and $elE_{z}=92$ meV. This gives $\omega_{1}>0$
and $\omega_{2}>0$, with $\left(\omega_{1}-\omega_{2}\right)\simeq3.6$
T, so we are plotting the region between the maximum at
$x_{M}=4/\left(\omega_{1}-\omega_{2}\right)$ and the minimum at $x_{m}=4.5/\left(\omega_{1}-\omega_{2}\right)$.
The temperatures considered, for this particular case, are specifically
chosen to represent how they affect the observation of the spin splitting (SP)
and the valley mixing (VM) behaviour that occurs around the beating minimum. Thus, starting from the $T=0$ case in
figure \ref{fig4}(a), we clearly observe the \emph{fine structure} of the MO,
due to the spin and valley. As we increase the temperature,
all the peaks start to broaden, and depending on $T$, some peaks
would no longer be observed. First of all, in figure \ref{fig4}(b), we see that
at $T=0.1$ K, the peaks are now broaden, but nevertheless one could
still appreciate the SP in the MO, which is seen as the
small bumps between the bigger peaks \cite{Escudero2017}. But if we continue
increasing the temperature, we get to the situation shown in figure \ref{fig4}(c),
where for $T=0.3$ K the MO are broaden such that the SP cannot be observed
any more. However, we still see the VM behaviour
in the MO when we are at the minimum region ($1/B$ around 1.25 T),
due to the broken valley degeneracy. Increasing the temperature further,
we eventually get to the state shown in figure \ref{fig4}(d), where the VM also disappears. This last state
is maintained when the temperature continues to increase, where the
MO are more damped but the form does not change, corresponding to a
pure beating phenomenon. We shall now study in detail the situations
considered, namely how we can estimate in general the temperature at which the SP and VM would no longer be observed. As discussed above, in this low temperature situation it will be more useful
to use equation (\ref{MOT}). In order to do that we will first analyse
how the last term in equation (\ref{MOT}) alters the observation
of each MO peak at non zero temperature.

\subsection{Temperature effect over each MO peak}

\begin{figure}[b]
	\includegraphics[scale=0.35]{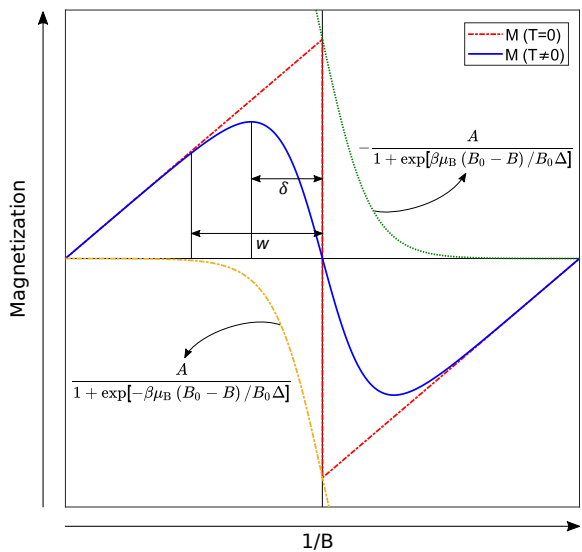}
	\caption{Schematic representation of the temperature effect over each MO peak,
		as expressed by equation (\ref{MOT}). In red (dashed line) it is shown
		a magnetization peak at $T=0$, located at $1/B_{0}$, and in blue
		(solid line) the magnetization at $T\protect\neq0$. It is also shown
		the exponentials that broaden the magnetization at non zero temperature,
		as described by equation (\ref{MT peak}). The peaks modification due
		to the temperature can be described by the parameters $w$ and $\delta$,
		as indicated. The width $w$ measures the reach of the temperature
		effect over each peak, while $\delta$ measures the extreme shift
		from its location at $T=0$.\label{fig5}}	
\end{figure}

We will study the temperature effect over each MO peak, which will
be useful in the subsequent analysis. Thus we consider, in general,
a unique MO peak at a given $B_{0}$ and we omit the effect of others
peaks close to $B_{0}$ (either due to spin splitting or valley mixing).
Then, from equation \ref{MOT} the magnetization is (we will take $s=1$, but
the analysis is valid for any spin
and valley)

\begin{equation}
M_{T}=\frac{A}{\pi}\arctan\left\{ \cot\left[\pi\omega\left(\frac{1}{B}+\Delta\right)+\pi \mathcal{F}_{0}\right]\right\},\label{MOT1}
\end{equation}
with $\mathcal{F}_{0}=\left\{ 1+\exp\left[\beta\mu_{\rm B}\left(B_{0}-B\right)/B_{0}\Delta\right]\right\} ^{-1}$.
The broadening of the MO at $T\neq0$ is entirely dictated by the
behaviour of $\mathcal{F}_{0}$. This can be seen considering that, by the properties of arctangent, equation (\ref{MOT1}) is
equivalent to \cite{Escudero2018a}

\begin{equation}
\label{cases}
M_T=\cases{M+A\left(1-\mathcal{F}_{0}\right) & $1/B<1/B_0$\\
	M-A\mathcal{F}_{0} & $1/B>1/B_{0}$\\},\label{MT peak}
\end{equation}
where $M=A\arctan\left\{ \cot\left[\pi\omega\left(\frac{1}{B}+\Delta\right)\right]\right\}/\pi$ is the magnetization at zero temperature. Notice the change of sign in the exponential, which is consistent
with the limit $M_{T}\rightarrow M$ if $T\rightarrow0$ ($\beta\rightarrow\infty$).
In figure \ref{fig5} it is schematically plotted $M$ and $M_{T}$, as a
function of $1/B,$ plus the exponential functions which give the temperature
correction. From this we can identify two properties of the broadening
due to the temperature: the shift $\delta$ of the extrema and the
width $w$ from which $M_{T}\simeq M$. Both $\delta$ and $w$ depend
on the temperature, and in general also depend on $B_{0}$ and $\mu$. They are obtained from the equations 

\begin{eqnarray}
\frac{\partial M_{T}}{\partial B}\left(\frac{1}{B}=\frac{1}{B_{0}}-\delta\right) & = & 0\\
\mathcal{F}_{0}\left(\frac{1}{B}=\frac{1}{B_{0}}-w\right) & \ll & 1\label{width}
\end{eqnarray}
The first equation can only be solved numerically.
In this way one obtains $\delta=\delta\left(T,B_{0}\right)$, and
in general, for the same temperature, $\delta\left(T,B_{01}\right)\neq\delta\left(T,B_{02}\right)$.
This dependence of $\delta$ with the magnetic field
implies a broken periodicity of the MO with $1/B$ at very low temperatures, although
usually one has $\delta\left(T,B_{01}\right)-\delta\left(T,B_{02}\right)\ll1/\omega$.
Nevertheless, as we will show later, when the temperature is increased
one needs to consider the effect of the surrounding peaks, in which
case the shift reaches the limit $\delta\rightarrow1/4\omega$, equal
to the medium of the maxima and zero of the MO peaks. On the other hand, the width
$w$ can be estimated from equation (\ref{width}) by choosing a cutoff
$\sigma\ll1$ such that $\mathcal{F}_{0}\left(1/B_{0}-w\right)=\sigma$. This gives
a width $w\simeq\ln\left(1/\sigma-1\right)k_{\rm B}T\Delta/\mu_{\rm B}B_{0}$,
where from our experience it is sufficient to take $\sigma\sim10^{-2}$
so $\ln\left(1/\sigma-1\right)\sim5$. The width $w$ is a measurement
of the local influence of the temperature over each MO peak, and as such it will be fundamental in estimating the temperature limits
corresponding to the different behaviours shown in figure \ref{fig4}.

\subsection{Spin splitting}

To study the influence of temperature over the observation of the
spin splitting (SP) in the MO, we follow the same lines as we did
in the graphene case \cite{Escudero2018a}, applying it to each of the frequencies now present.
Then we consider two MO peaks with frequency $\omega$, at a given
LL $n$, separated due to the SP, located in general at $1/B_{1}=n/\omega-\Delta$
and $1/B_{2}=n/\omega+\Delta$. From equation (\ref{MOT}), the corresponding
magnetization is

\begin{equation}
M_{T}=\frac{A}{\pi}\sum_{s=\pm1}\arctan\left\{ \cot\left[\pi\omega\left(\frac{1}{B}+s\Delta\right)+\pi \mathcal{F}_{12}\right]\right\} ,\label{MT SP}
\end{equation}
with $\mathcal{F}_{12}=\sum_{n=1,2}\left\{ 1+\exp\left[\beta\mu_{\rm B}\left(B_{n}-B\right)/\Delta B_{n}\right]\right\} ^{-1}$.
We know, from figure \ref{fig5}, that the width $w$ of this exponentials
determines the observation of the MO at non zero temperature. Thus, for
two peaks separated by $2\Delta$ due to the SP, one would expect
to see the SP in the MO only if $w<2\Delta$. In fact, this result
can be easily visualized by plotting equation (\ref{MT SP}) and the
corresponding exponentials, as done in the figure \ref{fig5}. This was done in
graphene \cite{Escudero2018a}, where one observes that as $w$ approaches $2\Delta$,
the SP disappears in the MO, and one is left with one oscillation
around the middle of the peaks. Consequently, given that $w\simeq5k_{\rm B}T\Delta/\mu_{\rm B}B$,
from the condition $w=2\Delta$ we get the spin temperature

\begin{equation}
T_{s}\simeq\frac{2\mu_{\rm B}B}{5k_{\rm B}},\label{Ts}
\end{equation}
where $B=\omega/n$ is the middle of two peaks separated due to the
SP. The condition to observe the spin splitting in the MO
is that $T<T_{s}$, which in order of magnitude means that the thermal
energy $k_{\rm B}T$ is lower than the Zeeman energy $2\mu_{\rm B}B$. This is the same temperature that was found in graphene,
which is expected because $T_{s}$ depends only on the spin splitting
effect in the MO and not on the broken valley degeneracy that appears
in 2D buckled crystals. Moreover, it does not depend on the particular
2D material properties, such as $\upsilon_{\rm F}$, $l$ or $\lambda_{\rm SO}$,
which again is expected because the SP alters the energy levels by
the introduction of the crystal independent Zeeman term $2\mu_{\rm B}B$. In the particular case considered in figure \ref{fig4}, we get that for $1/B\sim1.1$
1/T, we have $T_{s}\sim0.25$ K, so for the region of magnetic fields
considered, one would not observe the SP in the MO at $T>0.25$ K.
This is consistent with Figs. 4(b) and 4(c), where at $T=0.1\:\mathrm{K}<T_{s}$
we see the SP, but at $T=0.1\:\mathrm{K}>T_{s}$ we do not.

\subsection{Valley mixing}

\begin{figure}[b]
	\includegraphics[scale=0.25]{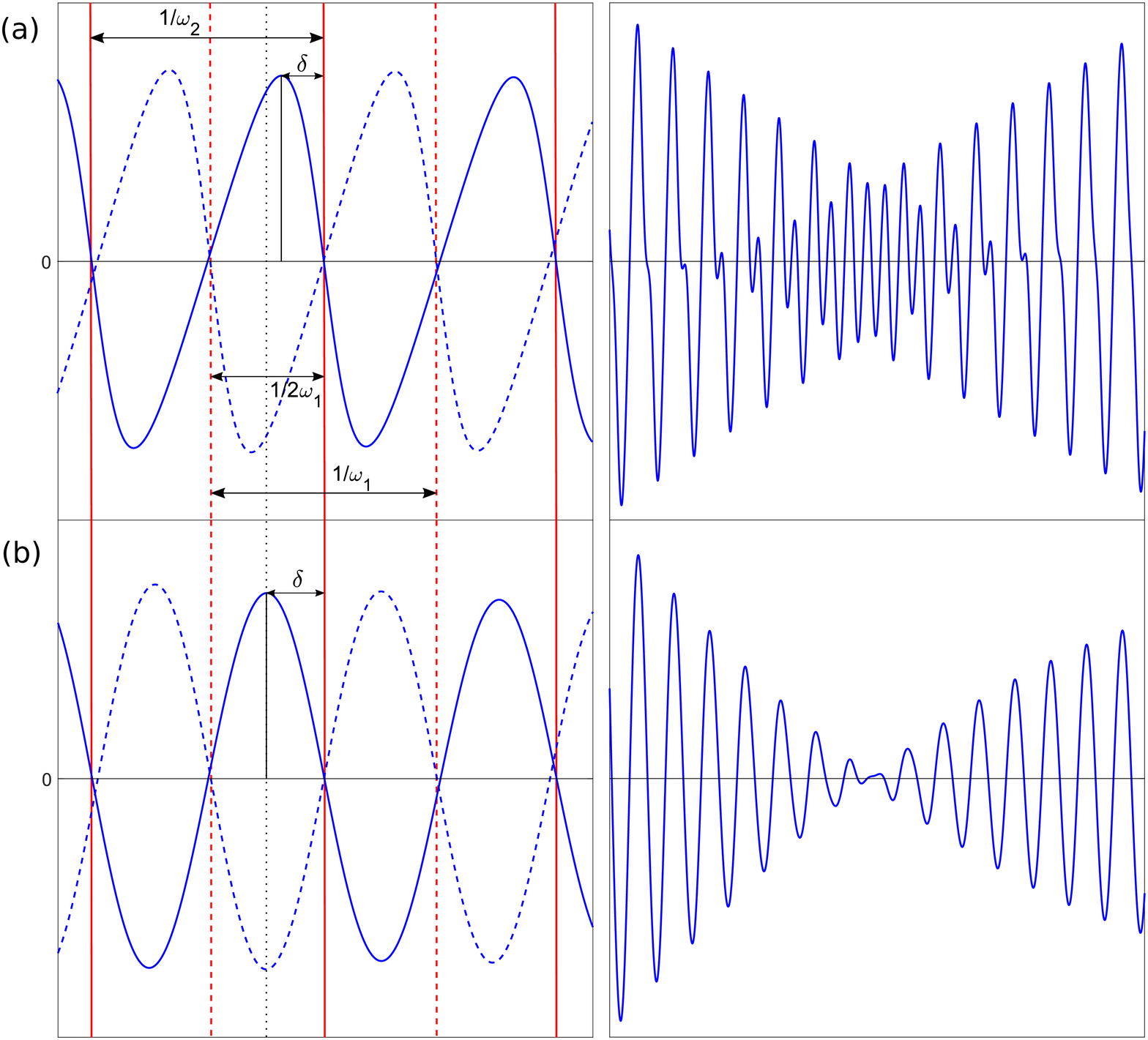}	
	\caption{Relationship between the valley mixing (VM) in the MO around the beating
		minimum, and the extrema shift $\delta$ of the MO peaks. On the left
		it is shown two MO peaks $M_{1}$ (dashed line) and $M_{2}$ (solid
		line), with frequencies $\omega_{1}$ and $\omega_{2}$, as a function
		of $1/B$, where the vertical lines in red correspond to the peaks
		location at $T=0$ while the blue lines are the MO at $T\protect\neq0$.
		The region of the plot corresponds to the minimum location in the
		MO, where there is destructive interference between the peaks. In
		the right it is shown the resulting MO around that minimum location,
		obtained by the sum of the $M_{1}$ and $M_{2}$. The case in (a)
		corresponds to a shift $\delta<1/4\omega$, for which the minimum and
		maximum of $M_{1}$ and $M_{2}$ are not in the same location, causing
		the mixing pattern seen in the resulting MO on the right. On the other
		hand, when the shift reaches the limit $\delta=1/4\omega$ in (b), the minimum
		and maximum of $M_{1}$ and $M_{2}$ are approximately in the same
		location, in which case there is no VM in the MO around the beating
		minimum.\label{fig6}}	
\end{figure}
We call the valley mixing (VM) effect in the MO to the mixing pattern
that appears around the MO minimum, as can be seen in figures \ref{fig4}(a)-(c).
This effect is independent of the SP and is caused by the broken valley
degeneracy, which overlaps the MO peaks with different frequency.
The way this overlap is produced determines how is the resulting
mixing behaviour. To understand this, consider the minimum condition,
when the peaks with frequency $\omega_{1}$ and $\omega_{2}$ are
between one another (destructive interference), as separated as possible.
This is shown in figure \ref{fig6}, where on the left we plotted the peaks with
frequency $\omega_{1}$ (dashed line) and $\omega_{2}$ (solid line),
and on the right the resulting MO obtained by their summation (for
the sake of simplicity, we shall omit the SP of each peak, but the
result obtained is independent of it). On the left, the vertical
lines in red correspond to the peaks locations at $T=0$, with their
periodicity $1/\omega$ indicated for each case. The first situation,
figure \ref{fig6}(a), corresponds to the case where the MO show a VM behaviour
around the minimum, as can be seen in the right figure. 
This can be
explained by analysing how the resulting MO is obtained from the summation
of the peaks shown in the left figure. There we see that, at this
temperature, the MO extrema shift $\delta$ is less than $1/4\omega$
(black dot line), so the maximum and minimum are not in the same location,
and therefore the MO do not become zero. On the other hand, when
the temperature increases, the extrema shift $\delta$ reaches the
limit $1/4\omega$ for both peaks, as shown in figure \ref{fig6}(b), in
which case the maximum and minimum are approximately in the same location.
This behaviour is maintained if the temperature is further increased,
for the shift $\delta$ remains at $1/4\omega$ and the increasing
of the temperature only reduces the amplitude of the oscillations.
Therefore, the condition to observer the VM in the MO is that the
extrema shift $\delta$ is less than $1/4\omega$. 

\begin{figure}
	\includegraphics[scale=0.35]{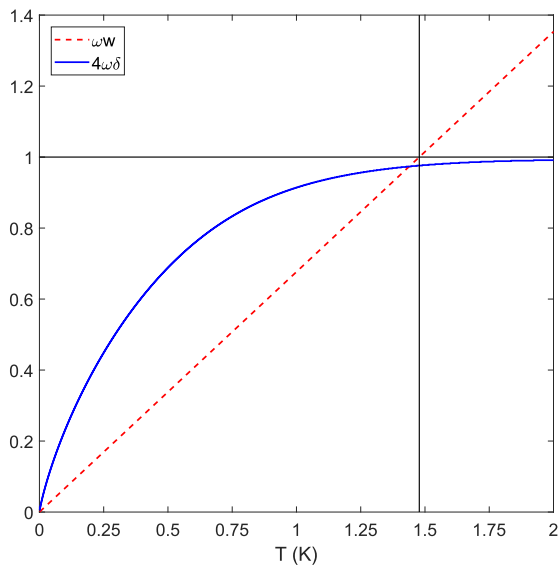}
	\caption{Relationship between the extreme shift $\delta$ and the width $w$
		of a MO peak, as defined in figure \ref{fig5}, for silicene, with $\mu=0.25$
		eV and $elE_{z}=92$ meV, considering the peak with frequency $\omega=\omega_{1}\simeq137.5$
		T at $1/B_{0}=172/\omega_{1}-\Delta_{1}=1.25$ 1/T. In the red dashed
		line it is plot $\omega w$, where $w=5T\Delta/\mu_{\rm B}B_{0}$ is the
		width of the exponential associated with the peak at $1/B_{0}$, as considered
		in equation (\ref{MOT}). In the blue solid line it is plot $4\omega\delta$,
		where $\delta$ is the maximum shift obtained numerically from 
		equation (\ref{shift}), with $B_{l}=B_{0}$ and $p=1$. As we see, $\delta$
		tends exponentially to $1/4\omega$, and when it does it we have $w\gtrsim1/\omega$.
		In other words, the maximum shift reaches its limit value when the
		range of the temperature influence over each MO peaks is bigger than
		the peaks separation $1/\omega$.\label{fig7}}	
\end{figure}

For the MO $M_{i,s}$, with frequency
$\omega_{i}$ and phase $s\Delta_{i}$, the extrema shift $\delta$
of a peak at $1/B_{l}=l/\omega_{i}-s\Delta_{i}$ is obtained from
the equation $\partial M_{i,s}/\partial B\left[1/B_{l}-p\delta_{p,l}\right]=0$,
where $p=1\:\left(-1\right)$ for the maximum (minimum) shift. Then, using  (\ref{MOT}), 
the equation for obtaining $\delta$ becomes 

\begin{equation}
1=\frac{\beta\alpha^{2}}{8\mu\left(1/B_{l}-p\delta_{p,l}\right)^{2}}\sum_{n}\frac{1}{B_{n}}\mathrm{sech^{2}}\left[\frac{\beta\alpha^{2}\omega_{i}}{4\mu}\left(\frac{1}{B_{l}}-\frac{1}{B_{n}}-p\delta_{p,l}\right)\right].\label{shift}
\end{equation}
The equation (\ref{shift}) can be solved numerically for each $B_{l}$ as a function
of the temperature, obtaining that $\delta$ follows an exponential
distribution, with the limit $\delta\rightarrow1/4\omega_{i}$. The
temperature at which we get this limit can be estimated from the relation of $\delta$ with the width
$w$ of the exponential associated with the peak at $B_{l}$, obtained
from equation (\ref{width}). This was done in figure \ref{fig7}, where we show the
numerical solution of equation (\ref{shift}) for $\delta$, and the width
$w=5k_{\rm B}T\Delta_{i}/\mu_{\rm B}B_{0}$ (considering $\mathcal{F}_0\left(1/B_{0}-w\right)=\sigma$
with $\sigma\sim10^{-2}$). 
The values correspond to silicene, with $\mu=0.25$ eV, $elE_{z}=92$
meV, and considering the spin up peak with frequency $\omega_{1}=137.53$
T at $l=172$, which gives $1/B_{l}\simeq1.25$ 1/T (thus the shift
and width calculated correspond to the peak around the minimum in
figure \ref{fig4}). Then we can see, in figure \ref{fig7}, that not only $\delta$
tends to the limit $1/4\omega_{1}$, but also that when it does it $w\gtrsim1/\omega_{1}$
. Hence, referring to figure \ref{fig5}, for a given peak the extrema shift $\delta$
approximately reaches its limit value when the width $w$ is about
the period of oscillation $1/\omega$. This gives an estimation
for the temperature $T_{v}$ at which $\delta\rightarrow1/4\omega$,
for then $w\sim1/\omega$ and therefore 

\begin{equation}
T_{v}\simeq\frac{\hbar\upsilon_{\rm F}^{2}eB}{5\mu k_{\rm B}}.\label{Tv}
\end{equation}
Then, following figure \ref{fig6},\emph{ }$T_{v}$ is also the temperature at
which the valley mixing would not longer be seen in the MO. In that
case it should be noted that the magnetic field that goes into 
equation (\ref{Tv}) corresponds to the peaks at the destructive interference,
or the absolute minima, that is $1/B=\left(r+1/2\right)/\left(\omega_{1}-\omega_{2}\right)$
with $r$ an integer. For the particular case of figure \ref{fig4} we get $T_{v}\simeq1.48$
K. This is in agreement with figure \ref{fig4}(d), where for $T=1.5$ K one
does not see the VM in the MO. It is interesting to compare the valley
temperature $T_{v}$ with the spin temperature $T_{s}$ given by equation
(\ref{Ts}). We have $T_{s}/T_{v}=2\mu_{\rm B}\mu/\hbar\upsilon_{\rm F}^{2}e$,
which from equations (\ref{Freq}) and (\ref{Delta}) implies $T_{s}/T_{v}=2\Delta\omega$.
Thus the ratio between these two temperatures is equal to the
ratio between the period $1/\omega$ and phase difference $2\Delta$
of the peaks (and this ratio is equal for all peaks). Of course, this
is an expected result because each temperature was calculated from
the width given by equation (\ref{width}), with $w=2\Delta$ for $T_{s}$
and $w=1/\omega$ for $T_{v}$.

\subsection{High temperature MO approximation}

When $T>T_{v}$, we can say the \emph{fine structure} of the MO is
damped by the temperature, and one is left with oscillations whose
extrema, for each frequency, are always shifted $1/4\omega$ from
the peaks locations at $T=0$. In this situation it becomes more convenient
to describe the MO using the LK formula given by equation (\ref{MT LK}),
for then we can approximate $\sinh\left(\lambda p\right)\sim\exp\left(-\lambda p\right)$.
Indeed, this approximation implies $\exp\left(-\lambda p\right)\ll1$
or $\lambda p\gg1$ ($\lambda=4\pi^{2}\mu k_{\rm B}T/\alpha^{2}B$), which
is satisfied for all $p$ if $T\gg\hbar\upsilon_{\rm F}^{2}eB/2\pi^{2}\mu k_{\rm B}=5T_{v}/2\pi^{2}\simeq T_{v}/4$.
Thus it is good approximation to take $\sinh\left(\lambda p\right)\sim\exp\left(-\lambda p\right)$
if $T>T_{v}.$ Then the summation over $p$ in equation (\ref{MT LK})
can be easily evaluated to $\lambda\sin\left[2\pi\omega_{i}\left(1/B+s\Delta_{i}\right)\right]\cosh\left(\lambda\right)$,
where we used the fact that in this regime $\cosh\left(\lambda\right)\gg1$
so $\cosh\left(\lambda\right)+\cos\left[2\pi\omega_{i}\left(1/B+s\Delta_{i}\right)\right]\simeq\cosh\left(\lambda\right)$.
We can further approximate the expression for $M_{T}$ by noticing
that in this \emph{high} temperature regime, the difference between
the amplitudes $A_{i}$ is practically negligible, so we can use the
amplitude $A_{p}/4$, where $A_{p}=2\sum_{i=1,2}A_{i}\simeq2\left[\lambda_{\rm SO}^{2}+\left(elE_{z}\right)^{2}-\mu^{2}\right]/\phi\mu$.
Then, rewriting the sine summation in equation (\ref{MT LK}), we get the
result

\begin{eqnarray}
M_{T} & \simeq & A_{p}\frac{k_{\rm B}T\gamma}{B}\textrm{sech}\left(\frac{\pi k_{\rm B}T\gamma}{B}\right)\cos\left(\gamma\mu_{\rm B}\right)\nonumber\\
& \times & \sin\left[\frac{\pi\left(\omega_{1}+\omega_{2}\right)}{B}\right]\cos\left[\frac{\pi\left(\omega_{1}-\omega_{2}\right)}{B}\right],\label{MT high}
\end{eqnarray}
where we defined $\gamma\equiv2\pi\mu/\hbar\upsilon_{\rm F}^{2}e$. It
is instructive to analyse each term in equation (\ref{MT high}). The
temperature effect is entirely contained in the term $\left(k_{\rm B}T\gamma/B\right)\textrm{sech}\left(\pi k_{\rm B}T\gamma/B\right)$,
which as expected goes to zero as $T$ increases, and it acts by just
reducing the overall amplitude of the MO. In other words, in this
regime the temperature does not modify the shape of each MO peak,
which of course is expected, as we are at temperatures such that we
already reached the limit $\delta\rightarrow1/4\omega$ for all $B$
considered. The term $\cos\left(\gamma\mu_{\rm B}\right)$ is independent of
the magnetic field and contains the effect due to the SP, which then
only acts as a reduction factor in the MO amplitude. Lastly, the last
two trigonometric functions in equation (\ref{MT high}) give the MO profile.
Under the beating condition $\left(\omega_{1}-\omega_{2}\right)/\omega_{1}\ll1$,
the first term causes the internal, small period oscillations, whereas
the second acts as the envelope of the internal oscillations. This
separation between each contribution will be particular useful in
order to obtain the MO envelope.

\subsection{MO envelope}

\begin{figure}[b]
	\includegraphics[scale=0.4]{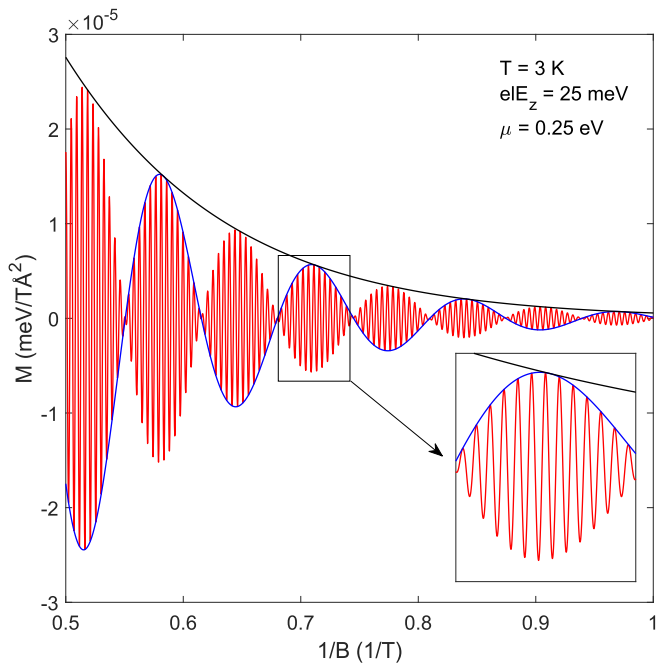}
	\caption{Magnetic oscillations in germanene, for $T=3$ K with $elE_{z}=25$
		meV and $\mu=0.25$ eV, such that $\left(\omega_{1}-\omega_{2}\right)/\omega_{1}\simeq0.07$.
		In this situation, the \textit{fine structure} of the MO is lost, given that
		from equation (\ref{Tv}) we have $T_{v}\simeq2.6$ K for $1/B=0.5$ 1/T. Thus the MO can
		be expressed with the Eq (\ref{MT high}). The beating envelope $E$
		in blue is given by equation (\ref{Envelope}), while the temperature decay
		envelope $E_{d}$ in black is $E=E_{d}\cos\left[\pi\left(\omega_{1}-\omega_{2}\right)/B\right]$.\label{fig8}}		
\end{figure}

We shall now obtain an expression for the MO envelope, restricting
ourselves to the beating condition, so that $\left(\omega_{1}-\omega_{2}\right)/\omega_{1}\ll1$.
In the general case, at a given temperature one should numerically
obtain the shift $\delta$ as function of $B$, and from it construct
the MO envelope, as was done in graphene \cite{Escudero2018a}. The generalization
to 2D materials with broken valley degeneracy is trivially done by
taking into account the two frequencies involved and the resulting
beating phenomenon. For simplicity we will omit this transition region
and consider only the case of \emph{high} temperatures, such that
$\delta=1/4\omega$. This implies $T>T_{v}$ for all the magnetic
field considered, in which case it is convenient to work with 
equation (\ref{MT high}) for the MO. To obtain its envelope we just have
to eliminate the internal oscillations in the sine function by evaluating it
at its maximum value. Thus we get envelope

\begin{equation}
E\simeq A_{p}\frac{k_{\rm B}T\gamma}{B}\textrm{sech}\left(\frac{\pi k_{\rm B}T\gamma}{B}\right)\cos\left(\gamma\mu_{\rm B}\right)\cos\left[\frac{\pi\left(\omega_{1}-\omega_{2}\right)}{B}\right].\label{Envelope}
\end{equation}
The fact that the envelope is obtained when $\sin\left[\pi\left(\omega_{1}+\omega_{2}\right)/B\right]=1$
implies that the extreme shift for the internal oscillations is of
the form $1/B=2l/\left(\omega_{1}+\omega_{2}\right)-1/2\left(\omega_{1}+\omega_{2}\right)$,
which resembles $1/B=l/\omega-\delta$ with $\omega=\left(\omega_{1}+\omega_{2}\right)/2$
and $\delta=1/4\omega$. Thus we get our previous result that at this
regime the extreme shift is equal to $1/4\omega$, with the frequency
being the average between $\omega_{1}$ and $\omega_{2}$. From equation
(\ref{Envelope}) we can also get the temperature decay envelope $E_{d}=E/\cos\left[\pi\left(\omega_{1}-\omega_{2}\right)/B\right]$.
In figure \ref{fig8} it is shown the MO and its envelope in germanene, for $T=3$ K with
$elE_{z}=25$ meV and $\mu=0.25$ eV. 
In this case, from equation (\ref{Tv}) we have $T_{v}\simeq2.6$ K for $1/B=0.5$ 1/T. Hence $T>T_v$ for all the $B$ considered and therefore the MO and its envelope are given by equations (\ref{MT high}) and (\ref{Envelope}). 
It should be noted that due
to the temperature, not only the oscillation amplitude is damped,
but also there is a shift of each MO beating maximum from its location
at $T=0$. At zero temperature, this maximum occurs at $1/B_{M}=r/\left(\omega_{1}-\omega_{2}\right)$,
with $r$ an integer (see figure \ref{fig3}), but at non zero temperature, the new maximum
occurs when $\partial E/\partial B=0$, and because the temperature
decay $E_{d}$ depends on $B$, then its solution is no longer $1/B_{M}.$
This can be seen in the zoomed oscillations in figure \ref{fig8}, where the
decay envelope $E_{d}$ (calculated considered the maxima at $1/B_{M}=r/\left(\omega_{1}-\omega_{2}\right)$)
does not exactly pass over the MO extrema. On the other hand, the
zeros in the MO are fixed at $\left(r+1/2\right)/\left(\omega_{1}-\omega_{2}\right)$,
the same location of the beating minimum at $T=0$. From an experimental
point of view, this is an useful result because the distance between
the MO nodes will be always the beating period $1/\left(\omega_{1}-\omega_{2}\right)=\hbar\upsilon_{\rm F}^{2}/2\lambda_{\rm SO}lE_{z}$.
Thus, independently of the temperature, by measuring when the MO is
zero one can obtain information about the crystal parameters. 

\section{Conclusions}

We studied the magnetic oscillations (MO) in 2D materials with a honeycomb buckled structure, in the low energy approximation, where the electrons are described by a Dirac-like Hamiltonian. Examples of these system are silicene, germanene, stanene and phosphorene. We considered a perpendicular electric and magnetic field, taking into account the spin orbit interaction (SOI) and the Zeeman effect. For a constant positive Fermi energy, we showed that, at zero temperature, the MO can be decomposed as the sum of four sawtooth oscillations (SO), associated with the change of valley and spin in the last energy level occupied. The four SO consist of two unique frequencies, each one with two different phases due to the spin splitting. The frequencies depend on the crystals properties, as well as the Fermi energy and electric field, and the corresponding oscillation occurs only if they are positive. Hence, depending on the values of the Fermi energy and electric field, one can have only one frequency in the MO, or directly no oscillation. When both frequencies are present, the MO show an interference pattern. A beating phenomenon is seen only if the frequencies are close, which results in a rombo-like pattern in the MO at zero temperature. When the frequencies are further apart, the MO show a more disperse, non-beating pattern, where the behaviour depends specifically on the values of the frequencies. We studied the condition to observe a beating in the MO, obtaining that the lower the Fermi energy, the lower the perpendicular electric field needs to be. 

At non zero temperature, we considered the broadening of the MO using two different approaches. One was the Lifshitz-Kosevich (LK) formula that considers the temperature effect by the introduction of a reduction factor. The other approach, recently developed in graphene, considers the temperature effect by local corrections over each MO peak and thus is particular useful at low temperatures. Using this last approach we studied how the increase of the temperature alters the observation of the \textit{fine structure} of the MO, due to the valley and spin. We showed that this can be related to the width of the Fermi-Dirac like functions that modify each magnetization peak at non zero temperature. Specifically, we obtained that  
in order to observe the spin splitting (SP), the width must be lower than the MO phase difference. Likewise, in order to observe valley mixing (VM) effects in the MO, the width must be lower than the MO period. When the temperature is such that the SP and VM are no longer seen, then the MO is best described by the LK formula, for one can approximate and easily evaluate the series. We then obtained a simple expression for the MO, and its envelope, where one can clearly see how the different frequencies produce a beating phenomenon. 

The results obtained show unique properties in the MO in 2D materials. The interplay between the valley and spin, under a perpendicular electric field, gives rise to oscillations with different frequencies and phases, a behaviour not seen in conventional metals. Because of this, by studying the shape of the MO one can obtain information about the 2D materials parameters. For instance, we showed the presence of zeros in the MO when there is a beating phenomenon, and the temperature is high enough such that the SP and VM are not longer appreciable. This may be useful from an experimental point of view, for the location of these zeros depends exclusively on the perpendicular electric field and the crystals properties, such as the Fermi velocity, buckle height and SOI. Lastly, we want to remark that the results obtained correspond to the pristine case, where no effect of impurities is considered.
It is known that the impurities also broaden the MO, so one would expect a similar behaviour to the one described for the temperature. In particular, the higher the impurities concentration, the lower the temperature necessary to observe the fine structure of the MO. On the other hand, an in-plane electric field would also be expected to modulate the MO, as has been reported in graphene.  

\ack
This paper was partially supported by grants of CONICET (Argentina
National Research Council) and Universidad Nacional del Sur (UNS)
and by ANPCyT through PICT 2014-1351. Res. N 270/15. N: 2014-1351, and PIP 2014-2016. Res. N 5013/14. C\'odigo: 11220130100436CO research grant, as well as by SGCyT-UNS., J.
S. A. and P. J. are members of CONICET., F. E. acknowledge  research fellowship from this institution.

\appendix
\setcounter{section}{0}
\section{MO at zero temperature}

We will derive equation (\ref{MO}) for the magnetic oscillations
at $T=0$, for the general case of a 2D crystal with energy levels
$\varepsilon_{\zeta,n,\eta,s}=\zeta\left[\left(s\lambda_{\rm SO}-\eta elE_{z}\right)^{2}+\alpha^{2}nB\right]^{1/2}-s\mu_{\rm B}B$,
where $\zeta=\pm1$ for the conduction and valence bands, $\alpha=\upsilon_{\rm F}\sqrt{2\hbar e}$, $n=0,\:1,\,2,\ldots$
for the Landau level (LL) and $\eta,s=\pm1$ are the valley and spin indices.
Graphene is a special case, with $\lambda_{\rm SO}\simeq0$ and $l=0$,
and therefore the derivation of $M$ will follow an analogous procedure
to the one employed in \cite{Escudero2018a}. We shall repeat the
essential steps of this derivation just for completeness. 

We consider a constant Fermi energy $\mu>0$, such that at zero temperature
the valence band is full while the conduction band is partially filled.
We will note the conduction energy levels $\varepsilon_{m}=\left[\left(s_{m}\lambda_{\rm SO}-\eta_{m}elE_{z}\right)^{2}+\alpha^{2}n_{m}B\right]^{1/2}-s_{m}\mu_{\rm B}B$,
where we have introduced the decreasing energy sorting index $m=0,\:1,\,2,\ldots$.,
so $n_{m}$ gives the LL, $\eta_{m}$ the valley and $s_{m}$ the
spin for the $m$ position. At a given $\mu>0$, all energy levels
$m=0,\:1,\,2,\ldots,\,f$ are filled, where $f$ is such that $\varepsilon_{f}\leq\mu<\varepsilon_{f+1}$.
Then the grand potential at zero temperature is $\Omega=\Omega_{V}+\sum_{m=0}^{f}D\left(\varepsilon_{m}-\mu\right)$,
where $\Omega_{V}$ is the grand potential due to the filled valence
band. It is important to notice that the oscillation in $\Omega$
is caused only by the last term, due to the conduction band, associated
with the change in the last energy level as $B$ is changed. On the
other hand, the first term $\Omega_{V}$ makes a non-oscillatory contribution
since the valence band is always filled for $\mu>0$. Separating $\varepsilon_{m}=\varepsilon_{m}^{0}-\varepsilon_{m}\mu_{\rm B}B$,
with $\varepsilon_{m}^{0}=\left[\left(s_{m}\lambda_{\rm SO}-\eta_{m}elE_{z}\right)^{2}+\alpha^{2}n_{m}B\right]^{1/2}$,
we can write the conduction grand potential as 

\begin{equation}
\Omega_{C}=\Omega_{0}-BM_{P},\label{GP3}
\end{equation}
where $\Omega_{0}=\sum_{m=0}^{f}D\left(\varepsilon_{m}^{0}-\mu\right)$
and $M_{P}=\mu_{\rm B}D\sum_{m=0}^{f}s_{m}$ is the Pauli paramagnetism
associated with the spin population. The conduction magnetization
is given by $M_{C}=-\mathcal{A}^{-1}\left(\partial\Omega_{C}/\partial B\right)_{\mu}$,
where $\mathcal{A}$ is the sheet area. Deriving and regrouping we
get

\begin{equation}
M_{C}=-\frac{1}{2B}\left(3\frac{\Omega_{C}}{\mathcal{A}}+\rho\mu\right)+M'+\frac{1}{2}m_{P},\label{Mc}
\end{equation}
where $\rho=N/\mathcal{A}=\sum_{m=0}^{f}D/\mathcal{A}=A\left(f-1\right)/D$
is the density of conduction electrons, $m_{p}=M_{P}/\mathcal{A}$ and

\begin{equation}
M'=\frac{e}{2h}\sum_{m=0}^{f}\frac{\left(s_{m}\lambda_{\rm SO}-\eta_{m}elE_{z}\right)^{2}}{\varepsilon_{m}+s_{m}\mu_{\rm B}B}.\label{M'}
\end{equation}
It is worth
noting that, looking at equation (\ref{M'}), we see that $M'$ is related
to the SOI and the buckle height. Thus this contribution is zero in
graphene, while in the other Dirac crystals it becomes appreciable, especially
at large electric field. From equation (\ref{Mc}) we directly see that the MO have a sawtooth
oscillation (SO) produced whenever $\rho$, $M'$ or $m_{P}$ change
discontinuously, $\Omega_{C}$ being continuous always. The SO amplitude
$\Delta M$ is given by

\begin{equation}
\Delta M=-\frac{\mu}{2B}\Delta \rho+\Delta M'+\frac{1}{2}\Delta m_{P},\label{DeltaM}
\end{equation}
where each contribution $\Delta\rho$, $\Delta M'$ and $\Delta m_{P}$
is determined by the discontinuous change in the parameters $n_{f}$, $\eta_{f}$
and $s_{f}$ which define the last energy level occupied. The SO peaks occur at $B_{i}$ such that $\varepsilon_{f}(B_{i})=\mu$.
Therefore $\mu=\left[\left(s_{i}\lambda_{\rm SO}-\eta_{i}elE_{z}\right)^{2}+\alpha^{2}n_{i}B_{i}\right]^{1/2}-s_{i}\mu_{\rm B}B_{i}$,
and given that usually $\mu_{\rm B}B/\mu\ll1$, we obtain 

\begin{equation}
\frac{1}{B_{i}}=\frac{n_{i}\alpha^{2}-2s_{i}\mu\mu_{\rm B}}{\mu^{2}-\left(s_{i}\lambda_{\rm SO}-\eta_{i}elE_{z}\right)^{2}}.\label{1/Bi}
\end{equation}
From this we can consider four types of MO peaks, taking into account
the possible changes of LL, valley and spin. Each peak is associated
to a fixed valley and spin, with its oscillation being caused when
the LL changes by one. The period of oscillation is $\Delta(1/B)=1/B_{2}-1/B_{1}$,
with $\Delta n=n_{2}-n_{1}=1$, while $\eta=\eta_{1}=\eta_{2}$ and
$s=s_{1}=s_{1}$. Consequently, from equation (\ref{1/Bi}) we obtain the
period $\Delta(1/B)$ and frequency $\omega=\left[\Delta(1/B)\right]^{-1}$

\begin{equation}
\omega_{\eta s}=\frac{\mu^{2}-\left(s\lambda_{\rm SO}-\eta elE_{z}\right)^{2}}{\alpha^{2}}.\label{freq}
\end{equation}
Then we can write equation (\ref{1/Bi}) as $1/B_{\eta s}\left(n_{i}\right)=n_{i}/\omega_{\eta s}+\Delta_{\eta s}$,
where $\Delta_{\eta s}$ is the phase 

\begin{equation}
\Delta_{\eta s}=-\frac{2s_{i}\mu\mu_{\rm B}}{\mu^{2}-\left(s\lambda_{\rm SO}-\eta elE_{z}\right)^{2}}.\label{phase}
\end{equation}
From equations (\ref{freq}) and (\ref{phase}) we see that $\omega_{K\uparrow}=\omega_{K'\downarrow}$,
$\omega_{K\downarrow}=\omega_{K'\uparrow}$ while $\Delta_{K\uparrow}=-\Delta_{K'\downarrow}$,
$\Delta_{K'\uparrow}=-\Delta_{K\downarrow}$, so there are two unique
frequencies and phases. Under the conditions assumed, the peaks can
only occur if $\omega_{\eta s}>0$ in equation (\ref{freq}). Indeed, remember
that equation (\ref{freq}) was derived from equation (\ref{1/Bi}) considering
$B_{1}$ and $B_{2}$ such that $\varepsilon_{1}=\mu=\varepsilon_{2}$,
with $n_{2}>n_{1}$. Thus, if $\omega_{\eta s}<0$ then $\mu^{2}<\left(s\lambda_{\rm SO}-\eta elE_{z}\right)^{2}$,
which for $\mu_{\rm B}B/\mu\ll1$ implies $2\varepsilon_{i}s\mu_{\rm B}>\alpha^{2}n_{i}$.
Therefore we have $2\left(\varepsilon_{2}-\varepsilon_{1}\right)s\mu_{B}>\alpha^{2}\left(n_{2}-n_{1}\right)$,
but $\varepsilon_{2}-\varepsilon_{1}=0$, so $0>\alpha^{2}\left(n_{2}-n_{1}\right)$.
This result means $n_{2}<n_{1}$, in contradiction with the
initial assumption of $n_{2}>n_{1}.$

The peaks amplitude $A_{\eta s}$ is obtained from equation (\ref{DeltaM}). Suppose
the magnetic field is increased so the last sorted position $f$ changes
to $f-1$. For $\Delta \rho$ and $\Delta m_{P}$ we easily get $\Delta \rho=D/\mathcal{A}=B/\phi$
and $\Delta m_{P}=D\mu_{\rm B}s_{f}/\mathcal{A}=B\mu_{\rm B}s_{f}/\phi$.
For $\Delta M'$, when the change is produced we have $\varepsilon_{f}=\mu$,
so from equation (\ref{M'}) we get $\Delta M'=\left(s_{f}\lambda_{\rm SO}-\eta_{f}elE_{z}\right)^{2}/2\phi\left(\mu+s_{f}\mu_{\rm B}B\right)$.
Thus in general

\begin{eqnarray}
A_{\eta s} & = & \frac{e}{2h}\left[\frac{\left(s\lambda_{\rm SO}-\eta elE_{z}\right)^{2}}{\mu+s\mu_{\rm B}B}-\mu+s\mu_{\rm B}B\right]\nonumber\label{amp}\\
& \simeq & -\frac{e}{2h}\frac{\alpha^2\omega_{\eta s}}{\mu},
\end{eqnarray}
where we consider $\mu_{\rm B}B/\mu\ll1$. We
are now in position to express the four SO, whose amplitude, frequency
and phase are obtained from equations (\ref{freq}), (\ref{phase}), (\ref{amp}).
Each type of peak can be expressed as an infinite series, so the SO
are written as

\begin{equation}
M_{saw}=\sum_{\eta s}A_{\eta s}\sum_{p=1}^{\infty}\frac{1}{\pi p}\sin\left[2\pi p\omega_{\eta s}\left(\frac{1}{B}-\Delta_{\eta s}\right)\right],\label{M sawtooth}
\end{equation}
The equation (\ref{M sawtooth}) gives the SO contribution to the MO. There
is still another oscillatory contribution, which comes from the continuous
oscillation in $\Omega_{C}$. From equation (\ref{M sawtooth}) we see
that $\Omega_{C}^{osc}$ should be of the form $\Omega_{C}^{osc}=\sum_{\eta s}C_{\eta s}\sum_{p=1}^{\infty}\cos\left[2\pi p\omega_{\eta s}\left(\frac{1}{B}-\Delta_{\eta s}\right)\right]/\left(\pi p\right)^{2},$
where $C_{\eta s}$ is such that $M_{osc}=-\mathcal{A}^{-1}\left(\partial\Omega_{C}^{osc}/\partial B\right)_{\mu}$.
From equation (\ref{M sawtooth}) we get $C_{\eta s}=-\mathcal{A}B^{2}A_{\eta s}/2\omega_{\eta s}$,
so the MO are given by

\begin{eqnarray}
M_{osc} & = & \sum_{\eta s}A_{\eta s}\sum_{p=1}^{\infty}\frac{1}{\pi p}\sin\left[2\pi p\omega_{\eta s}\left(\frac{1}{B}-\Delta_{\eta s}\right)\right]\nonumber\\
& - & \sum_{\eta s}A_{\eta s}\frac{B}{\omega_{\eta s}}\sum_{p=1}^{\infty}\frac{1}{\left(\pi p\right)^{2}}\cos\left[2\pi p\omega_{\eta s}\left(\frac{1}{B}-\Delta_{\eta s}\right)\right],\label{Mosc}
\end{eqnarray}
where we used the fact that $\partial A_{\eta s}/\partial B\simeq0$
so $\partial C_{\eta s}/\partial B\simeq-\mathcal{A}BA_{\eta s}/\omega_{\eta s}$.
The equation (\ref{Mosc}) is in agreement with \cite{Tabert2015}, where
the oscillating part of the magnetization is written as an infinite
series. There is still the non-oscillatory contribution to the magnetization,
from both the valence and conducting band. Nevertheless, it can be
shown \cite{Sharapov2004} that when $\mu>\left|s\lambda_{\rm SO}-\eta elE_{z}\right|$,
this contribution cancels and the total magnetization is given by
equation (\ref{Mosc}). From equation (\ref{freq}) we see that the condition
$\mu>\left|s\lambda_{\rm SO}-\eta elE_{z}\right|$ implies $\omega_{\eta s}>0$,
which as discussed above is also the condition to observe the oscillation
corresponding to the peak $\eta s$. Thus, because we will be mainly
interested with the MO, we shall omit the non-oscillatory contribution
and take the total magnetization given by equation (\ref{Mosc}). It is
worth noting that in this formalism the spin splitting due to the
Zeeman effect is already taken into account in equation (\ref{M sawtooth}),
so there is no need to introduce it as a reduction factor. 

We can further simplify equation (\ref{Mosc}) by noticing that the cosine
series is usually much smaller than the sine series. This can be seen
by analysing the corresponding series amplitude ratio, given by $A_{\eta s}^{cos}/A_{\eta s}^{sin}=B/\omega_{\eta s}$.
Considering that for all the 2D crystals we have $\alpha\sim10\:\mathrm{meV/\sqrt{T}}$,
and we will work with values around $\mu\sim10^{2}$ meV and $\left|s\lambda_{\rm SO}-\eta elE_{z}\right|\sim10$
meV, we have $A_{\eta s}^{cos}/A_{\eta s}^{sin}\sim10^{-2}B\left[\mathrm{T}\right]$.
Therefore, unless $B$ is very high we can neglect the cosine series
in equation \ref{Mosc}. Then the sine series can be easily evaluated to
obtain the MO at zero temperature

\begin{equation}
M=\sum_{\eta s}\frac{A_{\eta s}}{\pi}\arctan\left\{ \cot\left[\pi\omega_{\eta s}\left(\frac{1}{B}-\Delta_{\eta s}\right)\right]\right\} .\label{M0}
\end{equation}
Finally, equation  (\ref{M0}) can be conveniently rewritten by separating
the peaks with frequency $\omega_{1}=\omega_{K\uparrow}=\omega_{K'\downarrow}$
and $\omega_{2}=\omega_{K\downarrow}=\omega_{K'\uparrow}$, with phases
$\Delta_{1}=\Delta_{K\uparrow}=-\Delta_{K'\downarrow}$ and $\Delta_{2}=\Delta_{K'\uparrow}=-\Delta_{K\downarrow}$,
which leads to  equation (\ref{MO}) for the MO at zero temperature. 

\setcounter{section}{1}
\section{MO at non zero temperature}

We will derive the expression (\ref{MOT}) for the MO at non zero
temperature. As it was done in the zero temperature case, the
derivation will follow the same approach already applied in graphene
\cite{Escudero2018a}, which we shall repeat here for completeness. We start with
the grand potential $\Omega_{T}$ at $T\neq0$, for which we can use
its non-relativistic expression in the absence of impurities \cite{Sharapov2004,Tabert2014}.
It is convenient to separate $\Omega_{T}$ by the contribution
of each peak associated to the resulting MO. In other words, we separate
$\Omega_{T}=$$\sum_{i=1,2,s=\pm1}\Omega_{i,s}$, where

\begin{equation}
\Omega_{i,s}=-k_{\rm B}T\int_{-\infty}^{\infty}\rho_{i,s}\left(E\right)\ln\left[1+e^{\beta\left(\mu-E\right)}\right]dE.\label{OmegaT}
\end{equation}
Here $\beta=1/k_{\rm B}T$ and $\rho_{i,s}\left(E\right)=D\sum_{\zeta,n}\delta\left(E-\varepsilon_{i,s}\right)$
is the density of states (DOS) in the pristine case, where $\varepsilon_{i,s}=\zeta\left[\mu^{2}+\alpha^{2}\left(nB-\omega_{i}\right)\right]^{1/2}-s\mu_{\rm B}B$
are the corresponding energy levels (we omit the $\zeta$ and $n$
subscripts for simplicity), associated with the MO peaks with amplitude
$A_{i}$, frequency $\omega_{i}$ and phase $s\Delta_{i}$ given
by equations (\ref{Amp})-(\ref{Delta}), with $\zeta=\pm1$ for the
valence band (VB) and conduction band (CB). Replacing $\rho_{i,s}\left(E\right)$,
the equation (\ref{OmegaT}) becomes $\Omega_{i}=-k_{\rm B}TD\sum_{\zeta,n}\ln\left\{ 1+\exp\left[\beta\left(\mu-\varepsilon_{i,s}\right)\right]\right\} $.
The magnetization is given by $M_{T}=\sum_{i=1,2,s=\pm1}M_{i,s}$,
where $M_{i,s}=-\mathcal{A}^{-1}\left(\partial\Omega_{i,s}/\partial B\right)_{\mu}$.
Now, under the condition $\mu>0$ and low temperatures such that $\beta\mu\gg1$,
we always have $\beta\left(\mu-\varepsilon_{i,s}\right)\gg1$ for
the VB, so $\Omega_{i,s}^{V}\left(T\right)\simeq\Omega_{i,s}^{V}\left(T=0\right)$
and $\left(\partial\Omega_{i,s}^{V}/\partial B\right)\left(T\right)\simeq\left(\partial\Omega_{i,s}^{V}/\partial B\right)\left(T=0\right)$.
On the other hand, for the CB we get

\begin{eqnarray}
M_{i,s}^{C}=-\frac{1}{\mathcal{A}}\left(\frac{\partial\Omega_{i,s}^{C}}{\partial B}\right)_{\mu} & = & \frac{k_{\rm B}TD}{\mathcal{A}B}\sum_{n}\ln\left[1+e^{\beta\left(\mu-\varepsilon_{i,s}\right)}\right]\nonumber\\
& + & \frac{D}{\mathcal{A}}\sum_{n}\frac{\partial\varepsilon_{i,s}}{\partial B}\frac{1}{1+e^{-\beta\left(\mu-\varepsilon_{i,s}\right)}}.\label{partial Omega}
\end{eqnarray}
If $\mu$ is such that $\varepsilon_{i,s}\left(n=f\right)\leq\mu<\varepsilon_{i,s}\left(n=f+1\right)$,
then we can write equation (\ref{partial Omega}) as

\begin{eqnarray}
M_{i,s}^{C} & = & M_{i,s}^{C}\left(T=0\right)+\frac{k_{\rm B}TD}{\mathcal{A}B}\sum_{n=0}^{f}\ln\left[1+e^{-\beta\left(\mu-\varepsilon_{i,s}\right)}\right]\nonumber\\
& + & \frac{k_{\rm B}TD}{\mathcal{A}B}\sum_{n=f+1}^{\infty}\ln\left[1+e^{\beta\left(\mu-\varepsilon_{i,s}\right)}\right]+\frac{D}{\mathcal{A}}\sum_{n=0}^{f}\frac{\partial\varepsilon_{i,s}}{\partial B}\frac{1}{1+e^{\beta\left(\mu-\varepsilon_{i,s}\right)}}\nonumber\\
& - & \frac{D}{\mathcal{A}}\sum_{n=f+1}^{\infty}\frac{\partial\varepsilon_{i,s}}{\partial B}\frac{1}{1+e^{-\beta\left(\mu-\varepsilon_{i,s}\right)}},\label{partial Omega2}
\end{eqnarray}
where $M_{i,s}^{C}\left(T=0\right)=D\sum_{n=0}^{f}\left[\left(\mu-\varepsilon_{i,s}\right)/B-\partial\varepsilon_{i,s}/\partial B\right]/\mathcal{A}$
is the CB magnetization at zero temperature. Because we are considering
$\beta\mu\gg1$, then each term in the exponential is appreciable
only for $B$ such that $\varepsilon_{i,s}\left(B\right)\sim\mu$.
Hence, for each term $\beta\left(\mu-\varepsilon_{i,s}\right)$, we
can expand $\varepsilon_{i,s}$ around $B_{n}$, where $1/B_{n}=n/\omega_{i}-s\Delta_{i}$
and $\varepsilon_{i,s}\left(B_{n}\right)=\mu$. Thus $\left(\mu-\varepsilon_{i,s}\right)\simeq\mu_{\rm B}\left(B_{n}-B\right)/B_{n}\Delta_{i}$.
Furthermore, for the terms $\partial\varepsilon_{i,s}/\partial B$
it is good approximation to take

\begin{eqnarray}
\frac{\partial\varepsilon_{i,s}}{\partial B} & = & \frac{\left(\varepsilon_{i,s}\right)^{2}-\mu^{2}+\alpha^{2}\omega_{i}}{2B\left(\varepsilon_{i,s}+s\mu_{\rm B}B\right)}\nonumber\\
& \simeq & \frac{\alpha^{2}\omega_{i}}{2B\mu}=-\frac{A_{i}\mathcal{A}}{D}
\end{eqnarray}
where $A_{i,s}$ is given by equation (\ref{Amp}). From this we can also
see that the logarithmic terms in equation (\ref{partial Omega2})
are much smaller than the exponential terms, so we can neglect them.
Indeed, we always have $\ln\left\{ 1+\exp\left[\pm\beta\left(\mu-\varepsilon_{i,s}\right)\right]\right\} \leq\ln2<1$
for $n\leq f$ and $n\geq f+1$, while the ratio of amplitude between
both terms is $r\equiv\left|\frac{k_{\rm B}TD}{\mathcal{A}B}/A_{i}\right|=2k_{\rm B}T\mu/\alpha^{2}\omega_{i}$.
Then, given for the 2D crystals we have $\alpha\sim10\:\mathrm{meV/\sqrt{T}}$,
and we will work with values around $\mu\sim10^{2}$ meV and $\left|s\lambda_{\rm SO}-\eta elE_{z}\right|\sim10$
meV so $\omega_{i}\sim10^{2}$ T, we have $r\sim10^{-4}T\left[\mathrm{K}\right]$.
Thus, under the temperatures that we consider, is good approximation
to discard the logarithms terms in equation (\ref{partial Omega2}). In
this way, considering also the VB magnetization $M_{i,s}^{V}=M_{i,s}^{V}\left(T=0\right)$,
and summing over $i$ and $s$, we get the total magnetization 

\begin{eqnarray}
M_{T} & = & M-\sum_{i=1,2}A_{i}\sum_{s=\pm1}\left[\sum_{n=0}^{f}\frac{1}{1+e^{\beta\mu_{\rm B}\left(B_{n}-B\right)/B_{n}\Delta_{i}}}\right.\label{MT 1}\nonumber\\
& + & \left.\sum_{n=f+1}^{\infty}\frac{1}{1+e^{-\beta\mu_{\rm B}\left(B_{n}-B\right)/B_{n}\Delta_{i}}}\right],
\end{eqnarray}
where $M$ is the magnetization at zero temperature, given by
equation (\ref{MO}). The expression given by equation (\ref{MT 1}) holds under
the initial assumption $\varepsilon_{i,s}\left(n=f\right)\leq\mu<\varepsilon_{i,s}\left(n=f+1\right)$
for each $i,s$ peak, which in turn implies $1/B_{i,s}\left(n=f\right)\leq1/B<1/B_{i,s}\left(n=f+1\right)$.
Therefore, the temperature effect over the MO is to introduce factors
proportional to $\left\{ 1+\exp\left[\beta\mu_{\rm B}\left(B_{n}-B\right)/B_{n}\Delta_{i}\right]\right\} ^{-1}$
if $n\leq f$ and proportional to $\left\{ 1+\exp\left[\beta\mu_{\rm B}\left(B_{n}-B\right)/B_{n}\Delta_{i}\right]\right\} ^{-1}$
if $n>f$. Finally, from the properties of the arctangent and floor
functions, equation (\ref{MT 1}) can be generalized for all $B$ by
introducing the exponential factors inside the arctangent in $M$,
which leads to equation (\ref{MOT}). 

\section*{References}
\bibliography{references}	

\providecommand{\newblock}{}
\begin{thebibliography}{10}
\expandafter\ifx\csname url\endcsname\relax
  \def\url#1{{\tt #1}}\fi
\expandafter\ifx\csname urlprefix\endcsname\relax\def\urlprefix{URL }\fi
\providecommand{\eprint}[2][]{\url{#2}}

\bibitem{Novoselov2005}
Novoselov K~S, Geim A~K, Morozov S~V, Jiang D, Katsnelson M~I, Grigorieva I~V,
  Dubonos S~V and Firsov A~A 2005 {\em Nature\/} {\bf 438} 197--200

\bibitem{Geim2007}
Geim A~K and Novoselov K~S 2007 {\em Nature Materials\/} {\bf 6} 183--191

\bibitem{Zhang2005}
Zhang Y, Tan Y~W, Stormer H~L and Kim P 2005 {\em Nature\/} {\bf 438} 201--204

\bibitem{Mas-Balleste2011}
Mas-Ballest{\'{e}} R, G{\'{o}}mez-Navarro C, G{\'{o}}mez-Herrero J and Zamora F
  2011 {\em Nanoscale\/} {\bf 3} 20--30

\bibitem{Lin2016}
Lin Z {\em et~al.\/} 2016 {\em 2D Materials\/} {\bf 3} 042001

\bibitem{Gupta2015}
Gupta A, Sakthivel T and Seal S 2015 {\em Progress in Materials Science\/} {\bf
  73} 44--126

\bibitem{Zhao2016}
Zhao J {\em et~al.\/} 2016 {\em Progress in Materials Science\/} {\bf 83}
  24--151

\bibitem{Lay2015}
Lay G~L 2015 {\em Nature Nanotechnology\/} {\bf 10} 202--203

\bibitem{Zhuang2015}
Zhuang J, Xu X, Feng H, Li Z, Wang X and Du Y 2015 {\em Science Bulletin\/}
  {\bf 60} 1551--1562

\bibitem{Houssa2015}
Houssa M, Dimoulas A and Molle A 2015 {\em Journal of Physics: Condensed
  Matter\/} {\bf 27} 253002

\bibitem{Balendhran2014}
Balendhran S, Walia S, Nili H, Sriram S and Bhaskaran M 2014 {\em Small\/} {\bf
  11} 640--652

\bibitem{Davila2016}
D{\'{a}}vila M~E and Lay G~L 2016 {\em Scientific Reports\/} {\bf 6} 20714

\bibitem{Saxena2016}
Saxena S, Chaudhary R~P and Shukla S 2016 {\em Scientific Reports\/} {\bf 6}
  24182

\bibitem{Zhu2015}
feng Zhu F, jiong Chen W, Xu Y, lei Gao C, dan Guan D, hua Liu C, Qian D, Zhang
  S~C and feng Jia J 2015 {\em Nature Materials\/} {\bf 14} 1020--1025

\bibitem{Carvalho2016}
Carvalho A, Wang M, Zhu X, Rodin A~S, Su H and Neto A~H~C 2016 {\em Nature
  Reviews Materials\/} {\bf 1}

\bibitem{Cho2017}
Cho K, Yang J and Lu Y 2017 {\em Journal of Materials Research\/} {\bf 32}
  2839--2847

\bibitem{Liu2011}
Liu C~C, Jiang H and Yao Y 2011 {\em Physical Review B\/} {\bf 84} 195430

\bibitem{Spencer2016}
Spencer M~J and Morishita T (eds) 2016 {\em Silicene\/} (Springer International
  Publishing)

\bibitem{Neto2009}
Neto A~H~C, Guinea F, Peres N~M~R, Novoselov K~S and Geim A~K 2009 {\em Reviews
  of Modern Physics\/} {\bf 81} 109--162

\bibitem{Kou2017}
Kou L, Ma Y, Sun Z, Heine T and Chen C 2017 {\em The Journal of Physical
  Chemistry Letters\/} {\bf 8} 1905--1919

\bibitem{Tahir2013}
Tahir M and Schwingenschlögl U 2013 {\em Scientific Reports\/} {\bf 3} 1075

\bibitem{Ezawa2015}
Ezawa M 2015 {\em Journal of the Physical Society of Japan\/} {\bf 84} 121003

\bibitem{Huang2016}
Huang C, Zhou J, Wu H, Deng K, Jena P and Kan E 2016 {\em The Journal of
  Physical Chemistry Letters\/} {\bf 7} 1919--1924

\bibitem{Hsu2017}
Hsu C~H {\em et~al.\/} 2017 {\em Physical Review B\/} {\bf 96} 165426

\bibitem{Wang2016}
Wang H, Pi S~T, Kim J, Wang Z, Fu H~H and Wu R~Q 2016 {\em Physical Review B\/}
  {\bf 94} 035112

\bibitem{Zhang2016}
Zhang H, Zhou T, Zhang J, Zhao B, Yao Y and Yang Z 2016 {\em Physical Review
  B\/} {\bf 94} 235409

\bibitem{Ghazaryan2015}
Ghazaryan A and Chakraborty T 2015 {\em Physical Review B\/} {\bf 92} 165409

\bibitem{Ezawa2012}
Ezawa M 2012 {\em Journal of the Physical Society of Japan\/} {\bf 81} 064705

\bibitem{Liu2017}
Liu, Luo, Xu, Tian and Ren 2017 {\em Condensed Matter Physics\/} {\bf 20} 43701

\bibitem{Drummond2012}
Drummond N~D, Z{\'{o}}lyomi V and Fal'ko V~I 2012 {\em Physical Review B\/}
  {\bf 85} 075423

\bibitem{Du2014}
Du Y {\em et~al.\/} 2014 {\em {ACS} Nano\/} {\bf 8} 10019--10025

\bibitem{Aghaei2015}
Aghaei S~M and Calizo I 2015 {\em Journal of Applied Physics\/} {\bf 118}
  104304

\bibitem{Yan2015}
Yan J~A, Gao S~P, Stein R and Coard G 2015 {\em Physical Review B\/} {\bf 91}
  245403

\bibitem{Ni2011}
Ni Z, Liu Q, Tang K, Zheng J, Zhou J, Qin R, Gao Z, Yu D and Lu J 2011 {\em
  Nano Letters\/} {\bf 12} 113--118

\bibitem{Abbasi2018}
Abbasi A and Sardroodi J~J 2018 {\em Applied Surface Science\/} {\bf 456}
  290--301

\bibitem{Wang2017}
Wang T, Guo W, Wen L, Liu Y, Zhang B, Sheng K and Yin Y 2017 {\em Journal of
  Wuhan University of Technology-Mater. Sci. Ed.\/} {\bf 32} 213--216

\bibitem{Sharapov2004}
Sharapov S~G, Gusynin V~P and Beck H 2004 {\em Physical Review B\/} {\bf 69}
  075104

\bibitem{Tabert2015}
Tabert C~J, Carbotte J~P and Nicol E~J 2015 {\em Physical Review B\/} {\bf 91}
  035423

\bibitem{Hese2014}
He{\ss}e L and Richter K 2014 {\em Physical Review B\/} {\bf 90} 205424

\bibitem{Fu2011}
Fu Z~G, Wang Z~G, Li S~S and Zhang P 2011 {\em Chinese Physics B\/} {\bf 20}
  058103

\bibitem{Lukyanchuk2011}
Luk'yanchuk I~A 2011 {\em Low Temperature Physics\/} {\bf 37} 45--48

\bibitem{Uchoa2008}
Uchoa B, Kotov V~N, Peres N~M~R and Neto A~H~C 2008 {\em Physical Review
  Letters\/} {\bf 101} 026805

\bibitem{Ardenghi2015}
Ardenghi J~S, Bechthold P, Gonzalez E, Jasen P and Juan A 2015 {\em The
  European Physical Journal B\/} {\bf 88}

\bibitem{Ardenghi2014}
Ardenghi J, Bechthold P, Gonzalez E, Jasen P and Juan A 2014 {\em Physica B:
  Condensed Matter\/} {\bf 433} 28--36

\bibitem{Escudero2017a}
Escudero F, Sourrouille L, Ardenghi J and Jasen P 2017 {\em Superlattices and
  Microstructures\/} {\bf 101} 537--546

\bibitem{Escudero2018b}
Escudero F, Ardenghi J, Sourrouille L, Jasen P and Juan A 2018 {\em
  Superlattices and Microstructures\/} {\bf 113} 291--300

\bibitem{Escudero2018}
Escudero F, Ardenghi J and Jasen P 2018 {\em Journal of Magnetism and Magnetic
  Materials\/} {\bf 454} 131--138

\bibitem{Shoenberg1984}
Shoenberg D 1984 {\em Magnetic oscillations in metals\/} (Cambridge University
  Press)

\bibitem{Escudero2017}
Escudero F, Ardenghi J, Sourrouille L and Jasen P 2017 {\em Journal of
  Magnetism and Magnetic Materials\/} {\bf 429} 294--298

\bibitem{Goerbig2011}
Goerbig M~O 2011 {\em Reviews of Modern Physics\/} {\bf 83} 1193--1243

\bibitem{Ardenghi2013}
Ardenghi J, Bechthold P, Jasen P, Gonzalez E and Nagel O 2013 {\em Physica B:
  Condensed Matter\/} {\bf 427} 97--105

\bibitem{Lukose2007}
Lukose V, Shankar R and Baskaran G 2007 {\em Physical Review Letters\/} {\bf
  98} 116802

\bibitem{Zhang2010}
Zhang S, Ma N and Zhang E 2010 {\em Journal of Physics: Condensed Matter\/}
  {\bf 22} 115302

\bibitem{Escudero2018a}
Escudero F, Ardenghi J~S and Jasen P 2018 {\em Journal of Physics: Condensed
  Matter\/} {\bf 30} 275803

\bibitem{Khalid1988}
Khalid M~A, Reinders P~H~P and Springford M 1988 {\em Journal of Physics F:
  Metal Physics\/} {\bf 18} 1949--1964

\bibitem{Engelsberg1970}
Engelsberg S and Simpson G 1970 {\em Physical Review B\/} {\bf 2} 1657--1665

\bibitem{Tabert2014}
Tabert C~J and Carbotte J~P 2014 {\em Journal of Physics: Condensed Matter\/}
  {\bf 27} 015008

\end{thebibliography}

\end{document}